\documentclass[referee]{ieice}
\usepackage{graphicx}
\usepackage{latexsym}
\usepackage{amsmath}
\usepackage{amssymb,bm}
\usepackage[varg]{txfonts}
\usepackage{here}
\usepackage{epsfig}
\usepackage{graphicx}
\usepackage{epstopdf}
\usepackage{verbatim} 
\usepackage{comment}
\usepackage{subfigure}
\DeclareMathOperator*{\argmax}{arg\,max}

\newcommand{\AmSLaTeX}{%
 $\mathcal A$\lower.4ex\hbox{$\!\mathcal M\!$}$\mathcal S$-\LaTeX}
\def\BibTeX{{\rmfamily B\kern-.05em
 \textsc{i\kern-.025em b}\kern-.08em
  T\kern-.1667em\lower.7ex\hbox{E}\kern-.125emX}}
\makeatletter
\def\tmpcite#1{\@ifundefined{b@#1}{\textbf{?}}{\csname b@#1\endcsname}}%
\makeatother

\field{B}
\title{Design and implementation of lighting control system using battery-less wireless human detection sensor networks}

\authorlist{
\authorentry[yutao@mobile.ee.titech.ac.jp]{Tao Yu}{s}{Titech}
\authorentry{Yusuke Kuki}{s}{OsakaU}
\authorentry{Gento Matsushita}{s}{OsakaU}
\authorentry{Daiki Maehara}{s}{Titech}
\authorentry{Seiichi Sampei}{f}{OsakaU}
\authorentry{Kei Sakaguchi}{m}{Titech}
}
\affiliate[Titech]{The authors are with Tokyo Institute of Technology.}
\affiliate[OsakaU]{The authors are with Osaka University.}

\begin{document}
\maketitle

\begin{summary}
Artificial lighting is responsible for a large portion of total energy consumption and has great potential for energy saving. 
This paper designs an LED light control algorithm based on users' localization using multiple battery-less binary human detection sensors. The proposed lighting control system focuses on reducing office lighting energy consumption and satisfying users' illumination requirement.
Most current lighting control systems use infrared human detection sensors, but the poor detection probability, especially for a static user, makes it difficult to realize comfortable and effective lighting control. To improve the detection probability of each sensor, we proposed to locate sensors as close to each user as possible by using a battery-less wireless sensor network, in which all sensors can be placed freely in the space with high energy stability.
We also proposed to use a multi-sensor-based user localization algorithm to capture user's position more accurately and realize fine lighting control which works even with static users.
The system is actually implemented in an indoor office environment in a pilot project. A verification experiment is conducted by measuring the practical illumination and power consumption. The performance agrees with design expectations. It shows that the proposed LED lighting control system reduces the energy consumption significantly, 57\% compared to the batch control scheme, and satisfies user's illumination requirement with $100\%$ probability.
\end{summary}


\section{Introduction}
It has become a crucial issue to make more efforts on the optimization of energy usage, because of the limited resources and the consequent increase in energy cost. A better energy usage results in fewer energy bills, less grid load and less environmental impact. Energy consumed by artificial lighting system constitutes a big portion of total energy consumption. In Japan, annual energy consumption of lighting amounts to 150.6 TWh, accounting for 16\% of national energy consumption \cite{lightpc}. Therefore, it will make a notable positive impact for energy conservation to make lighting energy usage more efficient and to reduce energy consumption.

User occupancy is one of the key factors of an energy-efficient lighting control system \cite{lightreview}. 
In conventional occupancy-based batch lighting control systems, when the presence of any user is detected in an area, a controller switches all or several corresponding lights on, and when the absence is detected for a given delay period, it switches off the lights. Experimental study shows that in such system 20-26\% energy can be saved compared with manual switching \cite{compareswitch}.

Occupancy sensors have been used for detecting user's presence or absence. In \cite{1,2,4}, user occupancy or position is detected using infrared sensors by dividing the space into partitions according to sensor coverages and then differentiating the single triggered sensor or the set of triggered sensors. In \cite{3}, multiple infrared sensors are integrated as a sensor node to increase the detection capacities. Ultrasonic sensors \cite{ultrasonic,5}, RFID \cite{rfid} and surveillance cameras \cite{image} are also employed to obtain more accurate information of user's occupancy. 
Past studies are inadequate for the application of intelligent lighting controls in three main aspects. 
First, in long-term uninterruptible surveillance, to avoid the frequent battery changes, sensors have to be located in the ceiling or on the wall near the power source. The placements are limited and far from users, which greatly weaken the detection accuracy. 
Second, for infrared-sensor-based schemes, there is a lack of consideration of its large inevitable detection errors especially for a static object, but in an indoor environment user quite probably stays somewhere. Thus the systems would suffer from the possible incorrect light switching operations when detection error happens, or user is out of the coverage. 
Third, for schemes using ultrasonic sensors, RFID and surveillance cameras, although more accurate information of user's occupancy can be obtained for lighting controls, their consumed power and price are relatively high, and they are more invasive and difficult to install in existing rooms, compared with infrared sensors. 
Therefore, to address the challenges, we design a localization scheme for lighting control by using multiple battery-less infrared sensors to detect user's occupancy and positions with low energy consumption and easy installation. 

\begin{figure*}[!t]
\centering
\includegraphics[height=9cm,width=16cm]{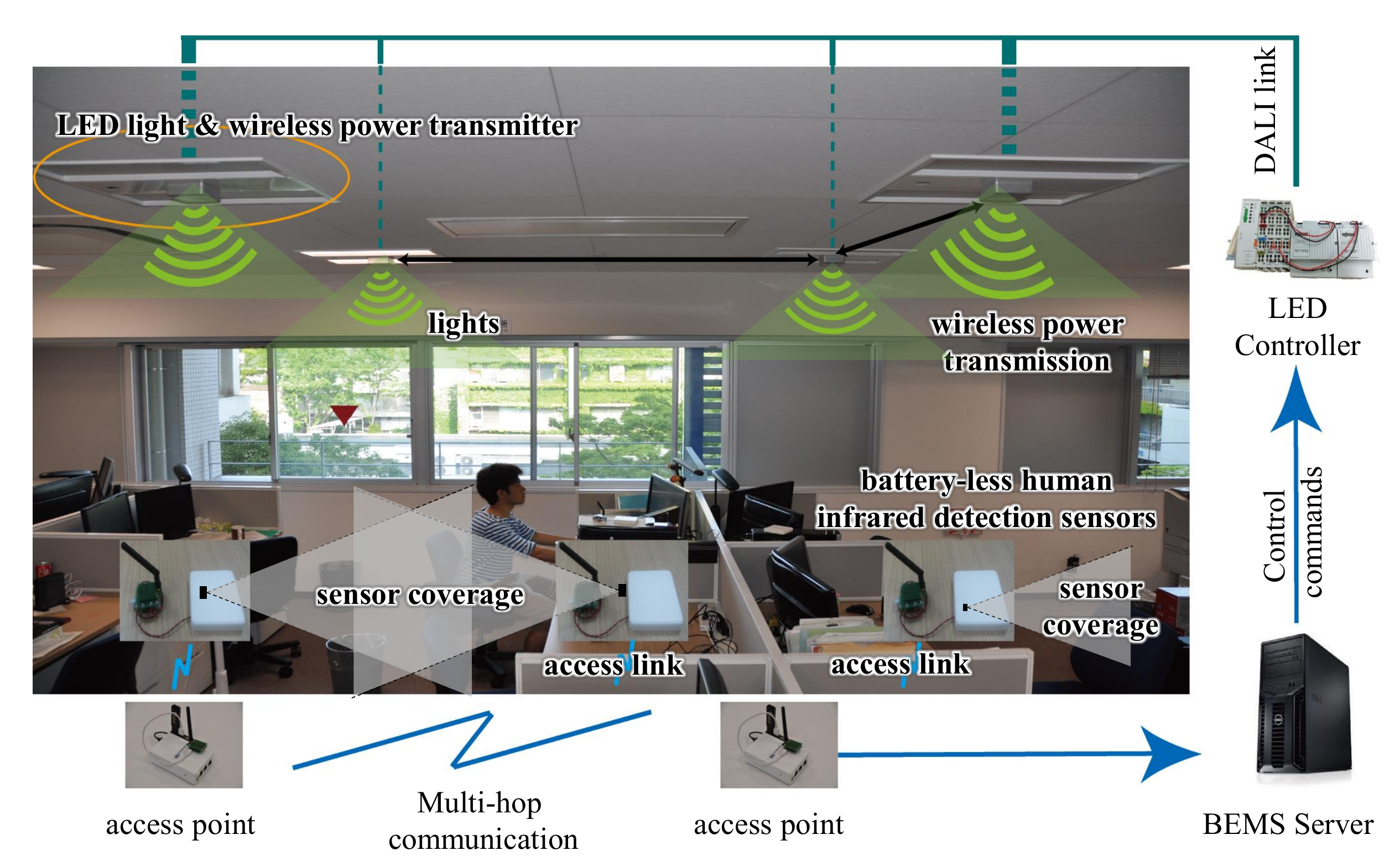}
\caption{Illustration of LED control system using battery-less sensor network.}
\label{fig:illustration}
\end{figure*}

Since the goal of the lighting control is to change the ON/OFF status or dimmer level of LED lights to create an area with convenient and productive illumination with minimum power consumption, 
the artificial lighting and environment illumination cannot be separately considered. 
Adaptively making the best use of environment illumination, such as daylight and reflected light, can highly decrease the requirement of artificial lighting. Therefore, combining both of them when designing a lighting project would efficiently increase the energy efficiency \cite{daylight,classroom,PaperR1,PaperR3,PaperR5,4}. 
Moreover, all research to date has detailed only simulations or small-scale testbeds. Our system has been implemented and applicated in a practical indoor office environment in E3-315, Osaka University, Japan \cite{yt}. The image of the system is illustrated in Fig.~\ref{fig:illustration}. 

In this paper, we design an LED lighting control scheme which bases on user's position and environment illumination level. It mainly focuses on the power consumption of the lighting system and the satisfaction of user's illumination requirement. It is suitable for office/home automation and can be easily installed in almost any environment without restriction. 
To deal with the poor detection probability of infrared sensors, we built a battery-less human detection network \cite{PaperMaehara}\cite{PaperMaehara2}, by which the sensors can be located anywhere in the space without restriction, so that they can be close to each user as much as possible, instead of being inflexibly embedded in walls or ceilings, which are far from detection objects.
And a series of distributed sensors with overlapping coverage were deployed, instead of single sensor for one LED light and one area in conventional lighting control systems. 
Based on the sensing measurements from multiple sensors and user's illumination requirement, the BEMS (Building Energy Management System) server estimates user's positions by a multi-sensor-based user localization algorithm, and generates control commands for switching ON/OFF or changing dimmer level of LED lights by solving an optimization problem in each control period. LED lights are set by LED controller. 

And a verification experiment is also conducted. In the experiment, the power consumption is logged by a power logger in office's electricity box, and the test user measures the practical illumination in his positions while walking around in the office. The experimental results show that the LED light control system can reduce power consumption by 57\% without any loss of user satisfaction.

The paper is organized as follows. 
Section~\ref{sec:system} describes the configuration and deployment of the LED light control system using a battery-less wireless human detection sensor networks. 
Section~\ref{sec:illumination_control} discusses the proposed LED light control algorithm.
Section~\ref{sec:localization} discusses user localization scheme. 
Section~\ref{sec:experiment} shows the results of a light control experiment in the office. 
Finally, Sec.~\ref{sec:conclusion} concludes this paper.

\section{System description}
\label{sec:system}

The overall image and structure of the LED light control system are illustrated in Fig.~\ref{fig:illustration}. It has been deployed in a practical indoor office environment in E3-315 of Osaka University. As shown in this figure, all LED lights are controlled by an LED light controller through a DALI (Digital Addressable Lighting Interface) link \cite{DALI1}\cite{DALI2}, in accordance with the commands generated by the BEMS server. And to capture user's location and to perform position-based light control, a battery-less wireless human detection sensor network is also implemented in the office \cite{PaperMaehara}\cite{PaperMaehara2}. Multiple battery-less infrared human detection sensors are placed on the desks with overlapped sensing coverages. The sensing measurements from sensors are transmitted to the BEMS server through multi-hop communication. The details of the system configuration and deployment will be described in this section.

\subsection{LED lights and controller}

\begin{figure}[!t]
	\centering
	\includegraphics[height=5.3cm,width=8.2cm]{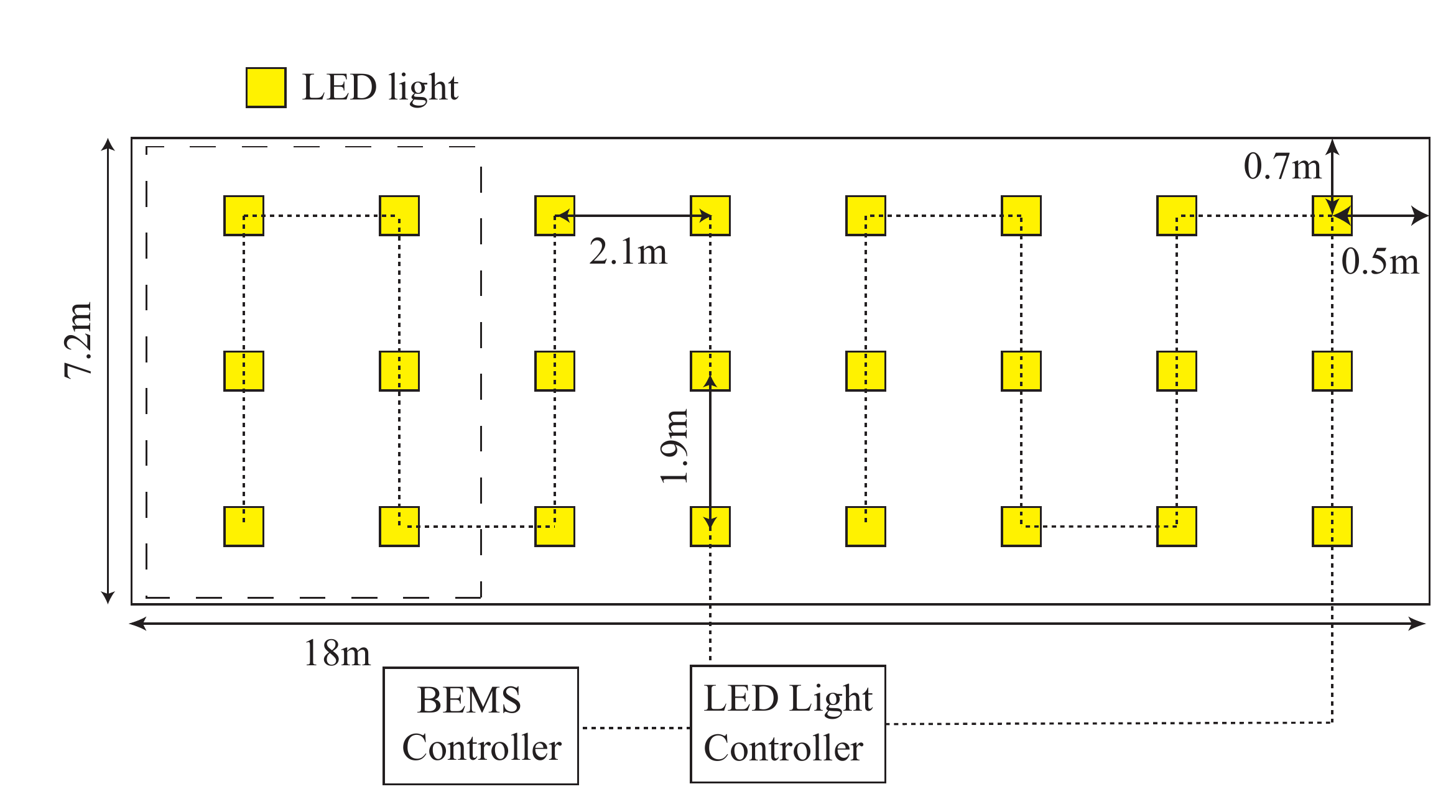}
	\caption{LED light and controller.}
	\label{fig:LED}
\end{figure}

LED lights are suitable for energy-efficient lighting control systems because compared to the fluorescent and incandescent light sources, LED lights have many merits including faster switching response time, lower energy consumption, higher physical robustness and longer lifespan.

In the target office of this work, shown in Fig.~\ref{fig:LED}, the LED lighting system consists of a plenty of LED light blocks which are spatially distributed on the ceiling of illuminated space. By adjusting the emitting illumination of the spatially distributed LED lights, it can offer more flexible illumination distribution and light rendering than conventional light sources. This adjustable illumination distribution can also save the energy spent on the illumination into areas where illumination is not actually required. 
 
All of the LED lights are connected to an LED light controller. The controller receives control commands from the BEMS server and then sends the commands to LED lights through a wired DALI link, which is a global industry standard protocol specifically designed for artificial lighting systems. The adoption of industrial standards simplifies the development and setup of such complex systems. Each DALI loop can control up to 64 devices each of which is fully addressable. The BEMS server can collect the sensing information from human detection sensors networks and create the LED control commands.

\subsection{Human detection sensor network}

\begin{figure}[!t]
	\centering
	\includegraphics[width=8.8cm]{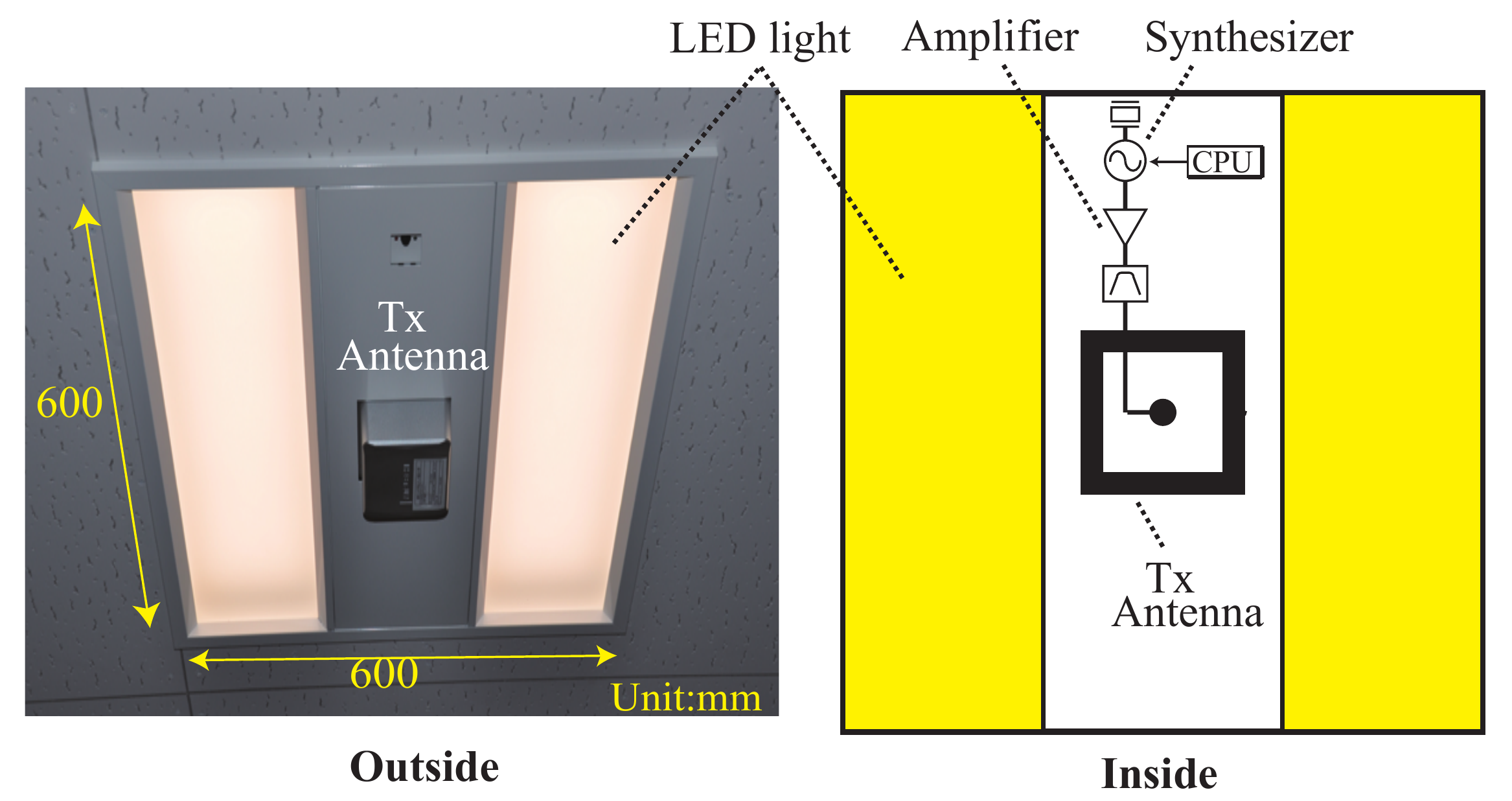}
	\caption{LED light and energy transmitter.}
	\label{fig:transmitter}
\end{figure}
\begin{figure}[!t]
	\centering
	\includegraphics[width=8cm]{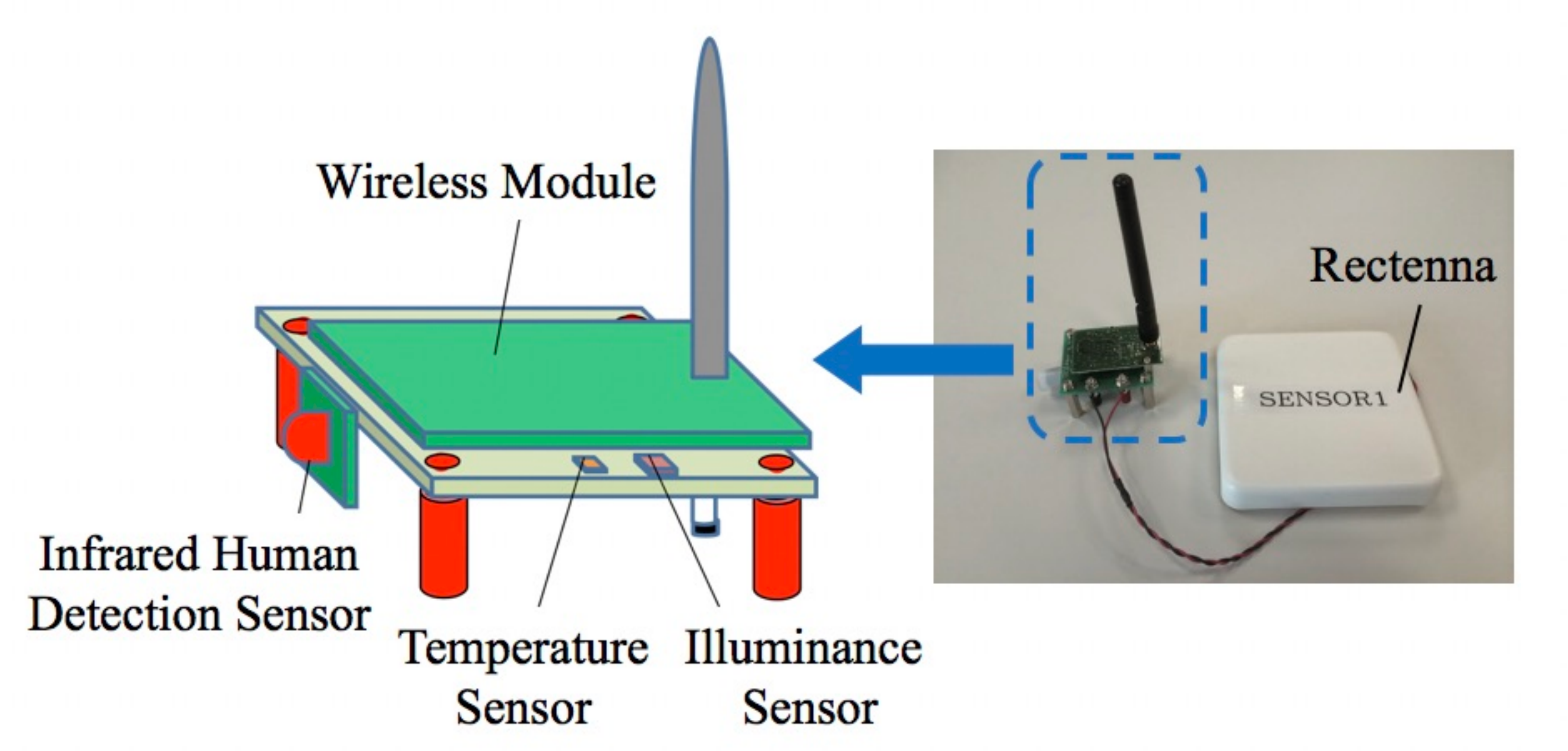}
	\caption{Battery-less sensors.}
	\label{fig:battery_less_sensor}
\end{figure} 

As is in most current conventional lighting control systems, binary infrared human detection sensor, which relies on detecting changes in the temperature pattern in the sensor's detection zone, is employed in this work, because its energy consumption and price are much lower than other types of human detection sensors (e.g., ultrasonic sensor and microwave sensor). The sensor of this work consumes only 150 $\mu$W in intermittent data transmission mode. 
The chief drawback of currently available infrared sensors is their low efficiency in detecting a static object and tiny movements. This makes it difficult to realize comfortable and effective lighting control.

To address the poor detection probability, instead of placing the sensors in the ceiling or on the walls, we proposed to more flexibly locate sensors close to each user as much as possible, by building a battery-less wireless sensor network in the target office, as shown in Fig.~\ref{fig:illustration}. 
That is to say, in this work, all sensors are battery-less and activated by multiple wireless energy transmitters, which are embedded in the ceiling LED lights, as shown in Fig.~\ref{fig:transmitter} and Fig.~\ref{fig:battery_less_sensor}. 
Because of wireless power transmission and wireless sensing data communication, sensors can be put anywhere in the room rather than such fixed locations as walls and ceilings. More sensors can be deployed in hot areas and at user's side for more accurate detection. 
Moreover, there is almost no need for follow-up maintenance, such as recharging or changing the batteries. 

On the transmitting side, the power of each energy transmitter is 1~W. Moreover, carrier shift diversity is employed \cite{PaperMaehara}\cite{PaperMaehara2}, by which the interference among multiple energy transmitters can be effectively avoided, and hence the continuous and seamless coverage of energy supply can be achieved in the office. 
The energy efficiency of power transmitter can be further increased in the future by using intermittent energy transmission, high efficient rectennas, beaming control, energy harvesting sensors, etc. \cite{energyhavesting,rectennas,beamcontrol} 

On the other hand, the sensor's function is simple. The binary infrared human detection sensor cannot directly give the information of user's location and movement. In each transmission process, it can only output 1-bit sensing information, which represents an estimation whether the user is present (1) or absent (0) in its coverage. To solve this problem, we proposed to deploy a distributed set of sensors with overlapping coverages in the human detection sensor network, instead of a single sensor for one LED light and one area.

And to sense the environment illumination, illumination sensors are embedded alone with each human detection sensor and also activated by wireless power transmission, as shown in Fig.~\ref{fig:battery_less_sensor}. Thus the illumination sensors are much closer to user's working surface and can be placed more flexibly, compared with the conventional illumination sensors which are located in the ceiling.

And on the receiving side, with the consideration of power conversion efficiency of rectenna, the sensors in a battery-less sensor node needs 400 $\mu$W to be activated battery-lessly.

\subsection{Multi-hop communication network}
The communication system is one of the core parts of any intelligent lighting control system. Wired communications for sensors will result in not only costly remodeling and rewiring work, but also inflexible sensor placement. In this work, sensors and BEMS server form the networks wirelessly, so that sensors can be easily deployed and integrated into the existing rooms without much restriction of installation. It is an accessible and economical solution for light control systems. To note, the communication between controller and light is through a wired DALI link.
 
Each sensor node detects user's presence/absence status, measures the illumination, and then sends the sensing data to an access point through a wireless channel. The available power for a sensor is limited because it is activated by wireless power transmission. Therefore, to guarantee the converge of the wireless sensor network without significantly increasing the power consumption of the RF module in each sensor node, it is sensible and necessary to introduce multi-hop communication network. Thus a hierarchical communication network is built, in which the access points have two wireless modules: sensing measurements collection, and multi-hop communication network among access points \cite{802} to create a backhaul network. 

Moreover, to reduce power consumption of sensors and extend the effective coverage of wireless power transmission, an intermittent data transmission scheme is employed. The sensor's sensing module, which is responsible for a very small amount of power compared to the RF module, is always active, but the RF module has two working modes (transmission mode/sleep mode), and intermittently transmits the output to the access points \cite{PaperMaehara2}.

\subsection{Implementation cost}
The LED lights and light controller (DALI) are all common retail productions and have no difference from those used in our daily life or in any other LED lighting control systems. The wireless power transmission system is also a general-purpose system and can continuously supply energy wirelessly to variety kinds of sensors and devices. We expect that in the future it would be located at ceilings or integrated into lights as a standard feature in every room \cite{PaperMaehara2}. 
Therefore, the only specialized part in this work is the wireless sensor network. It includes sensors, wireless modules, development board, USB interface, APs for sensing data transmission and APs for multi-hop communication. They totally cost around 60,000JPY in this work, and they could be much cheaper if mass produced.

\section{Illumination control}
The illumination control in this work mainly focuses on minimizing the lighting power consumption while satisfying user's illumination requirements, based on the user's position and illumination preference.
\label{sec:illumination_control}

\subsection{Consumed power of LED lights}
The power efficiency of LED lights, i.e., the ratio of consumed power to emitting illumination, is a function of many factors, which includes temperature, optical loss, working time, electrical and spectral efficiency, etc.~\cite{ledpower1}\cite{ledpower2} If the LED light is not overheated, its power efficiency can be controlled to be a constant level in stable environments, such as indoor facilities. Thus the relationship between emitting illumination and power consumption of LED lights can be considered as:
\begin{equation}
	C^l = {r^l}\,F^l
\end{equation}
where $C^l$ is the consumed power of LED light $l$, ${r^l}$ denotes the power efficiency factor of LED light $l$, and $F^l$ is the LED light's emitting illumination.

Therefore, the overall power consumption $C_{\rm{All}}$ of the LED light control system is given by summation of the power consumed by all LED lights and the other devices as:
\begin{equation}
	C_{\rm{All}} = C_{\rm{S}} + \sum_{l}^{I} {C^l} = C_{\rm{S}} + \sum_{l}^{I} {{r^l}\,F^l}
\end{equation}
where $I$ denotes the number of LED lights, and $C_{\rm{S}}$ is the system's overhead power consumption which includes the power consumed by the wireless power transmitter, wireless access point, BEMS server and LED light controller, etc. Note that compared to LED lights' power (maximum 100 W per LED light in this work), $C_{\rm{S}}$ is relatively small (around tens of watts). Therefore, it is treated as constant standby power here. Moreover, this part of energy consumption could be further decreased, e.g., by using lower energy cost devices and more efficient wireless energy transmission.

\subsection{Lighting model}

\begin{figure}[!t]
	\centering
	\includegraphics[width=8.8cm]{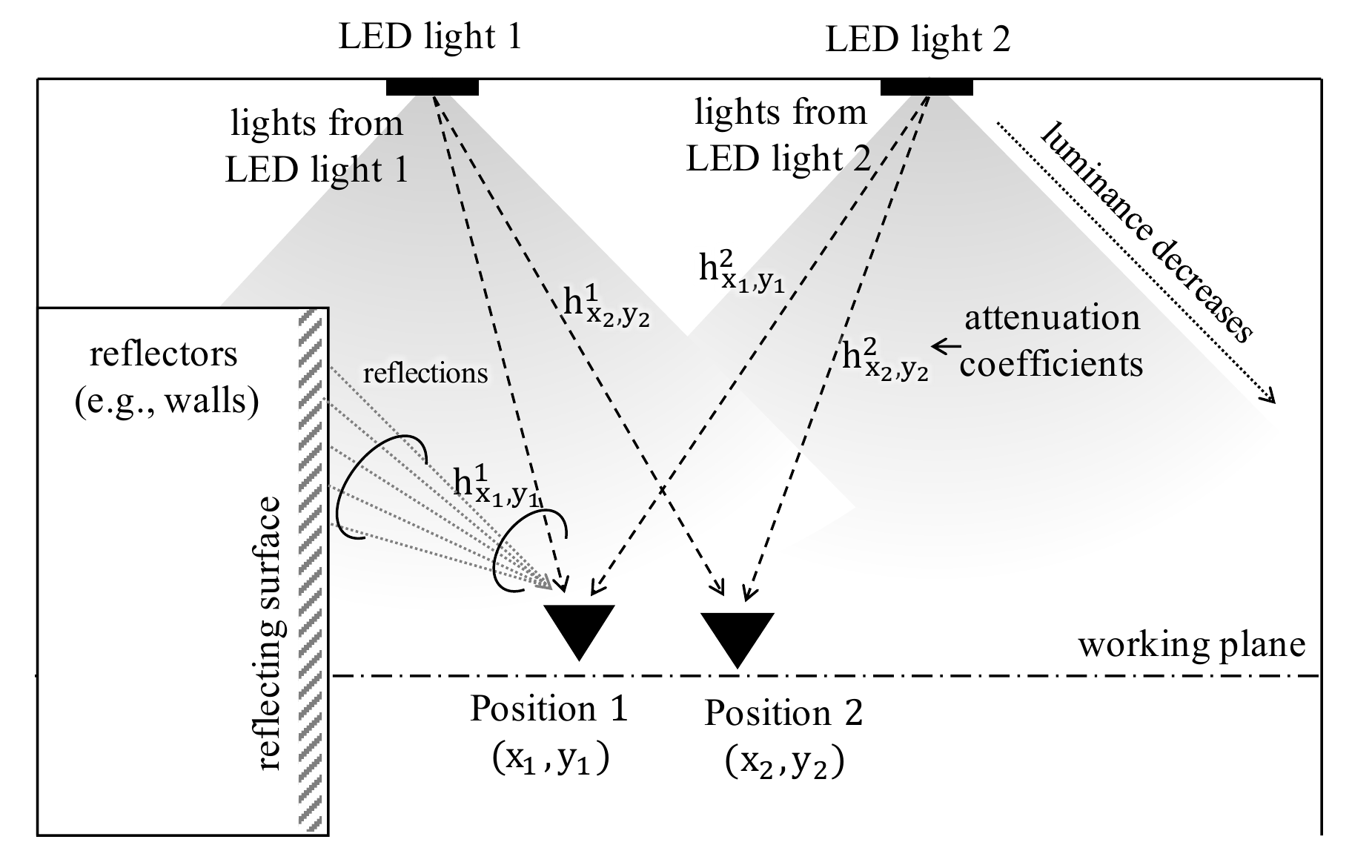}
	\caption{Lighting model.}
	\label{fig:lighting_model}
\end{figure}

Figure~\ref{fig:lighting_model} illustrates the geometry of LED lights in the ceiling and illuminated positions in working surface, which is a flat plane at desk height in the space. The illumination distribution of each position can be denoted by:
\begin{equation}
	\label{eq:illumination_distribution}
	F_{x,y}^l = F^l \, h_{x,y}^l = sw_l \,  F_{\rm{full}}^l \,  h_{x,y}^l
\end{equation}
where $F_{x,y}^l$ is the received illumination in position $(x,y)$ contributed by LED light $l$. $F^l = sw_l F_{\rm{full}}^l$ denotes luminous power of the LED light $l$, which is related to its maximum emitting illumination $F_{\rm{full}}$ and current switch status $sw_l$ of working dimmer level. $h_{x,y}^l$ is $l$th LED light's attenuation factor, which indicates the path loss in light's propagation from an LED light to an illuminated position. Obviously, the attenuation factor depends on the specification of LED lights (e.g. the physical shape and beaming angle of LED lights), the system geometry (e.g. the spatial relationship between LED lights and illuminated positions) and the target environment (e.g. building materials which result in light reflection). In most practical cases, such parameters will be nearly constant after setting up the system, so the factor $h_{x,y}^l$ can also be consequently considered as time-invariant.

Theoretically, the LOS (Line of Sight) propagation of lights from LED can be conveniently approximated by the generalized Lambertian source model, and $h_{x,y}^l$ can be accordingly defined by Lambertian functions following the Lambert's cosine law \cite{LambertianModel}. 
However practically, various common furnishing and building materials, such as walls and furniture surfaces, are efficient diffuse reflectors of visible light. In comparison with LOS propagation, it is much more complicated and cumbersome to analytically model the effect of multiple diffuse reflections by reflectors of different materials in the complex circumstance of a practical room. The illumination contributions from reflections are unneglectable and necessary to be taken into consideration for the total illumination of each illuminated position.
For example, the typical indoor smooth white wall can reflect almost 90\% of light. If there is a lack of such consideration of reflection, the system has to cost extra energy to satisfy the illumination requirement near reflectors, and this will consequently decrease the system’s energy saving performance. 
 
In the most cases, the dominated indoor reflectors, such as white walls and large smooth surfaces of furniture, are fixed after the construction of the room. Therefore, alternatively in this work, rather than model-based calculation, we practically obtain $h_{x,y}^l$ by respectively measuring $F'^l$ (the measured emitting illumination from LED light $l$) and $F'_{x,y}$ (the measured received illumination at illuminated position), while other LED lights are off, using illumination sensors:
\begin{equation}
 	h_{x,y}^l = F'_{x,y} \; / \; F'^l 
\end{equation}

\begin{figure}[!t]
	\centering
	\includegraphics[width=6cm]{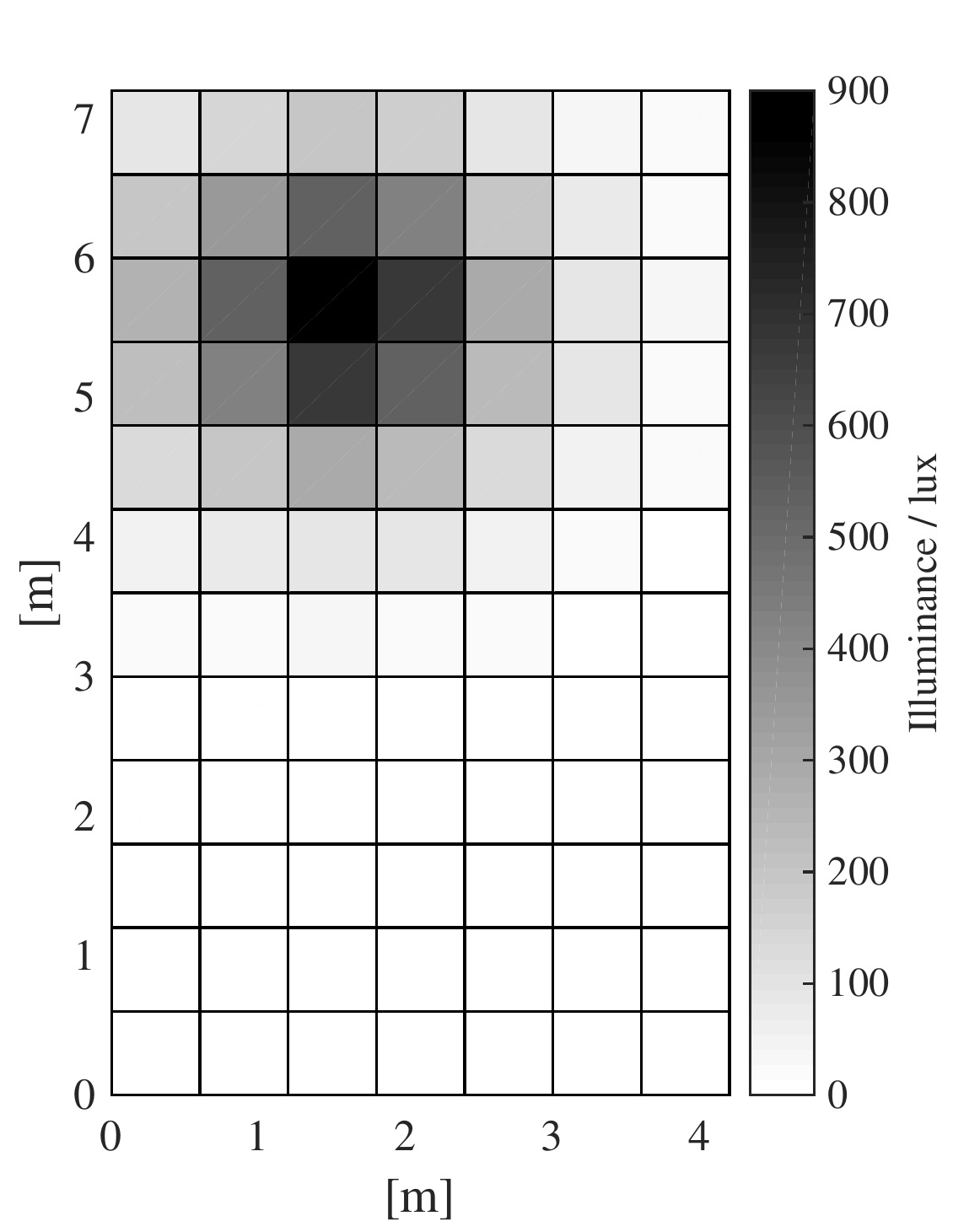}
	\caption{Measured illumination.}
	\label{fig:hlux}
\end{figure}

Figure~\ref{fig:hlux} gives an example of the measured received illumination from a full-on LED light in the left quarter of the office, which is shown in the dashed box in Fig.~\ref{fig:LED}. Dividing it by the LED light's maximum emitting illumination, the attenuation factor can be derived.
The benefits of such method are that both of direct illumination and the multiple diffuse reflections of lights in the practical room are well involved in total illumination without complicated environment and lighting model. 

In an illuminated position, its total illumination is contributed by the direct and reflected illumination from all LED lights. It is denoted as: 
\begin{equation}
	F_{x,y} = \sum_{l=1}^{I} F_{x,y}^l =  \sum_{l=1}^{I} sw_l F_{\rm{full}}^l h_{x,y}^l
\end{equation}

The environment illumination in each position is approximated by that of the closest illumination sensor, which is co-located beside the infrared human detection sensor and spatially distributed in the office.
\begin{equation}
	F_{x,y}^{\rm{en}} \approx F_{N(x,y)}^{\rm{en}}
\end{equation}
where $F_{x,y}^{\rm{en}}$ denotes the environment illumination in position $(x,y)$. $N(x,y)$ is the position of nearest illumination sensor to $(x,y)$.
\begin{eqnarray}
	\quad \quad\quad \quad F_{N(x,y)}^{\rm{en}} &=& F^s - F_{N(x,y)} \nonumber \\
						 &=& F^s - \sum_{l=1}^{I} sw_l F_{\rm{full}}^l h_{N(x,y)}^l
\end{eqnarray}
where $F^s$ is the illumination measurements from illumination sensors. $F_{N(x,y)}$ is the illumination contributed by LED lights.

\subsection{LED light control}

In an office, only the illuminations on and around occupants or some specified objects which need to be illuminated (e.g., a notice board) are necessary. And only when the necessary illuminations are higher than a satisfaction level, they are productive and can increase the efficiency at work and the health of people. Therefore, we consider a localized illumination control scheme.

From the analysis above, given the location of user and the satisfaction illumination level, the optimization problem of illumination control can be derived:
\begin{equation}
\label{eq:illumination_control}
	\begin{cases}

	\mathop {\min }\limits_{sw_1,sw_2...} C_{\rm{All}} = \sum\limits_l { {r^l} sw_l F_{\rm{full}}^l+C_{\rm{S}} } \\

	\quad \quad \quad s.t. \\

	F_{x,y} > F_{u;x,y}^{\rm{min}} - F_{x,y}^{\rm{en}}\quad (x,y) \in G_u \quad u = 1,2...U\\

	{0 \leq sw_l \leq 1 \quad l = 1,2...L}

	\end{cases}
\end{equation}
where $F_{x,y}^{\rm{en}}$ denotes the environment illumination on position $(x,y)$, which is obtained from illumination sensors. $F_{u;x,y}^{\rm{min}}$ is the minimum satisfaction illumination of user $u$, which can vary in different positions. $U$ is the number of users. Switch status $sw_l \in [0,1]$ indicates a certain dimmer level of LED light $l$.

Note that currently in the verification experiments only ON/OFF control is conducted, i.e., $sw_l = 1 \; or \; 0$. Therefore, some other considerations on illumination distribution (e.g., contrast ratio of illumination in an area \cite{humanfactorsinlighting} to increase users' satisfaction) are not implemented. The more flexible and convenient dimmer control scheme will be conducted soon. 

\section{User localization}
\label{sec:localization}
To realize the localized illumination control presented in Sec.~\ref{sec:illumination_control}, a battery-less wireless human detection sensor network is implemented to detect user's position and moving status. As described in Sec.~\ref{sec:system}, a series of battery-less infrared human detection sensors are distributed in the target room. The binary sensing measurements are used to estimate user's moving states.

\subsection{State transition model}

\begin{figure}[!t]
	\centering
	\includegraphics[width=8cm]{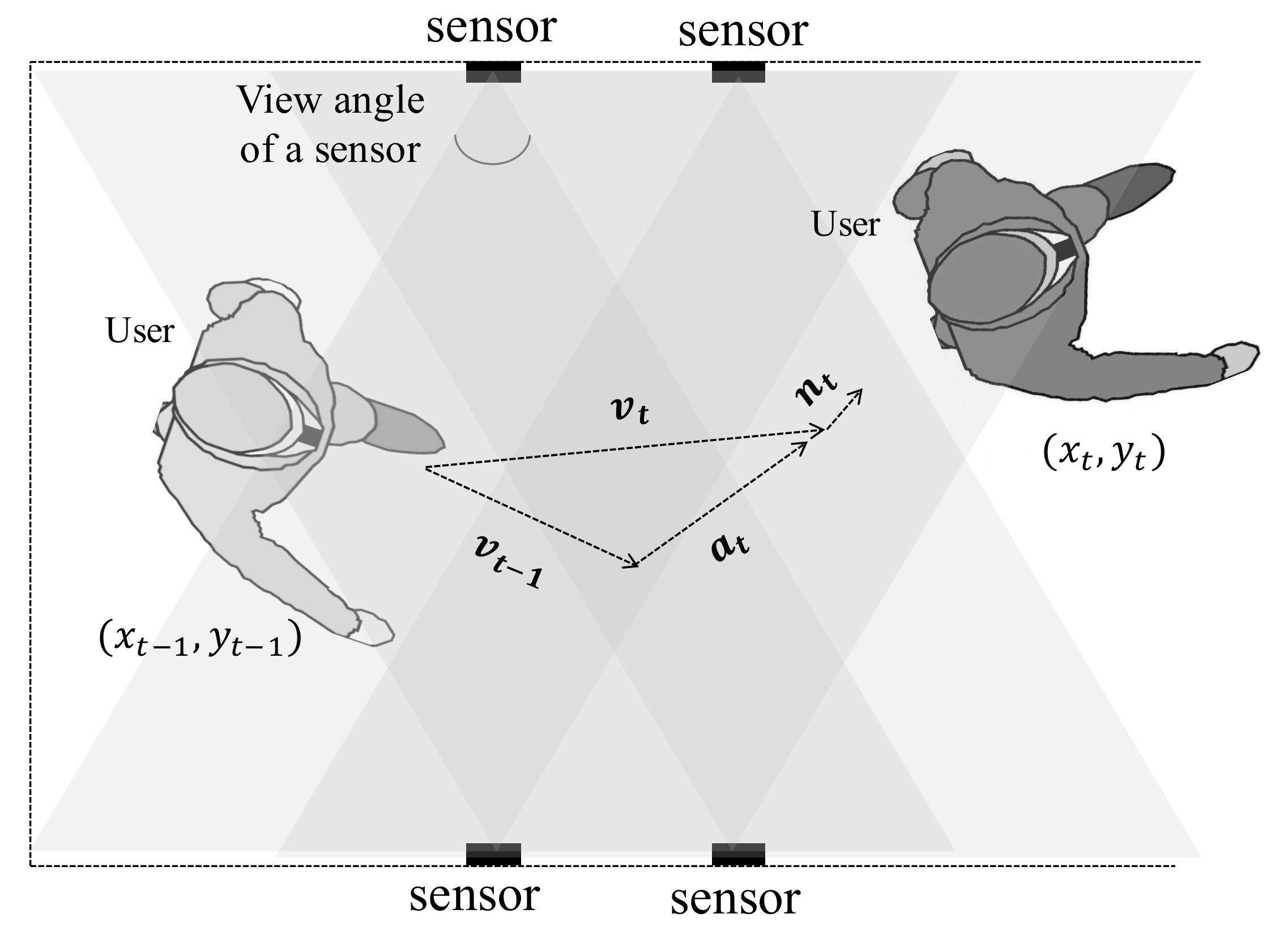}
	\caption{User moving model.}
	\label{fig:moving}
\end{figure}

In this paper we consider a 4-D state space $\bm{\mathcal{{S}}}\subset \bm{\mathcal{R}}^{4}$, with the state vector $\bm{s}=[ x,\; y,\; v_{x}\; ,v_{y} ]^{T}$, $\bm{s}\in \bm{\mathcal{S}}$, where $x$, $y$ and $v_{x}$, $v_{y}$ represent user's position and velocity in $x$-axis and $y$-axis respectively. 

The state transition model takes into account that in the target indoor environment, user's motion typically has two modes with different characteristics. \emph{Static Mode}: user stays in some location for a period. \emph{Moving Mode}: user walks following certain moving-pattern. Therefore, we use a switching-mode approach to model it based on the consideration above.

User's state transition rules in the two modes are different. 
In the moving mode, as shown in Fig.~\ref{fig:moving}, the states transition follows a variable acceleration motion. Normal distribution is adopted here as the model for the variable accelerations \cite{GaussianMotion}. Additionally, to model the non-uniform acceleration or non-rectilinear motion within a time interval, two normally distributed deviations are added to user's position in both $x$- and $y$-axis respectively. On the other hand, in the static mode, user's current state should either be the same as last state with probability $Pr^{\rm{S}}()$, or turn into moving mode with probability $1-Pr^{\rm{S}}()$. Naturally this mode should only contain the states with zero velocity. 
The self-transition probability $Pr^{\rm{S}}()$ is a predefined function of user's last state. For example, in an office environment, in the area near the desk, user is more likely to maintain the static mode with higher $Pr^{\rm{S}}()$. 
The transitions of other states will follow the moving model.

Therefore, the state transition model can be written as:
\begin{equation}
	\bm{u_t}= 
 	  \bm{u_{t-1}}+ ( \bm{A}\bm{u_{t-1}}+\bm{r_t}+\bm{n_t} ) \ I(\bm{u_{t-1}})
\end{equation}
with
\[\bm{A} = \left[ 
\setlength{\arraycolsep}{3pt}
{\begin{array}{*{20}{c}}
0&0&{\Delta t}&0\\
0&0&0&{\Delta t}\\
0&0&1&0\\
0&0&0&1
\end{array}} \right]\;,
\;\bm{r_t} = \left[ {\begin{array}{*{20}{c}}
{{a_{t,x}}\Delta {t^2}/2}\\
{{a_{t,y}}\Delta {t^2}/2}\\
{{a_{t,x}}\Delta t}\\
{{a_{t,y}}\Delta t}
\end{array}} \right]\;,
\;\bm{n_t} = \left[ {\begin{array}{*{20}{c}}
{ {n_{t,x}}}\\
{ {n_{t,y}}}\\
{0}\\
{0}
\end{array}} \right],
\]%
where, $\bm{u_{t}}=\bm{s} \in \bm{{S}}$ is user's state in time $t$. $\bm{A}$ is the state transition matrix for linear motion in moving mode. $\Delta {t}$ is the time interval. $a_{t,x},a_{t,y} \scriptsize{\sim} \mathcal{N}\left( 0,\sigma _{a}^{2} \right)$ are the independent normal accelerations in $x$-axis and $y$-axis. 
As described above, the deviation movement is modeled by normal velocities following $\mathcal{N}\left( 0,\sigma _{v}^{2} \right)$. Therefore, the deviations $n_{t,x},n_{t,y} \scriptsize{\sim} \mathcal{N}\left( 0,\sigma _{n}^{2} \right)$ with $\sigma _{n}=\sigma _{v}\Delta t$. 

$I(\bm{u_{t-1}})$ is an indicator function for user's moving mode, which is related to user's last state $\bm{u_{t-1}}$. It is given by:%
\begin{equation}
	I(\bm{u_{t-1}})=
	\begin{cases}
		\begin{array}{*{20}{cl}}
			\bm{0} & \text{ with Prob.}\   Pr^{\rm{S}}(\bm{u_{t-1}})\\
			\bm{1} & \text{ with Prob.}\ 1-Pr^{\rm{S}}(\bm{u_{t-1}}) 				
		\end{array}	
 	\end{cases}
\end{equation}

Therefore, the transition probabilities that user is in state $m$ in time $t-1$ and in state $j$ in time $t$ can be written as:
\begin{equation}
\label{eq:transition}
T_{ij}= 
	\begin{cases}
 		\ Pr^{\rm{S}}(\bm{s_i}) \text{\ \ \ \ \ \ \ \ \ \ \ \ \ \ \ \ \ \ \ \ \ \ \ \ when } i=j\\ 
		 \begin{split}
 			\ \!\![1\!-\!Pr^{\rm{S}}(\bm{s_i})]Pr^{\rm{M}}(\bm{u_t} = \bm{s_j}|\bm{u_{t - 1}} = \bm{s_i}) \\\text{ when } i\neq j
 		\end{split}
	\end{cases}
\end{equation}
with
\begin{equation*}
	Pr^{\rm{M}}(\bm{u_t} = \bm{s_j}|\bm{u_{t - 1}} = \bm{s_i},\ \bm{\xi _t},\ \bm{\Sigma_s})	\sim \mathcal{N}\left( \bm{\xi _{t}},\bm{\Sigma _{s}} \right)
\end{equation*}
where $\bm{\xi _t}$ and $\bm{\Sigma _s}$ are mean vector and covariance matrix respectively:
\begin{equation*}
	\bm{\xi _t} = [ 
		{{x_{t - 1}} + {v_{t - 1,x}}\Delta t},\ 
		{{y_{t - 1}} + {v_{t - 1,y}}\Delta t},\ 
		{{v_{t - 1,x}}},\ 
		{{v_{t - 1,y}}}
		]^T,
\end{equation*}
\begin{equation*}
	\bm{\Sigma _s} = 
	\left[ 
		{\begin{smallmatrix}
			{\sigma _v^2\Delta {t^2} + \frac{{\sigma _a^2\Delta {t^4}}}{4}}&0&{\frac{{\sigma _a^2\Delta {t^3}}}{2}}&0\\
			0&{\sigma _v^2\Delta {t^2} + \frac{{\sigma _a^2\Delta {t^4}}}{4}}&0&{\frac{{\sigma _a^2\Delta {t^3}}}{2}}\\
			{\frac{{\sigma _v^2\Delta {t^3}}}{2}}&0&{\sigma _a^2\Delta {t^2}}&0\\
			0&{\frac{{\sigma _v^2\Delta {t^3}}}{2}}&0&{\sigma _a^2\Delta {t^2}}
		\end{smallmatrix}} 
	\right]
\end{equation*}
For convenience, $\bm{\xi _t}$ and $\bm{\Sigma_s}$ are left out in the following content from the probability equation.

In the practical case, the moving model should be modified according to the real environment. The transition probabilities from or to invalid positions, e.g., positions in the invalid area, such as furniture's location, or out of the edges of the target area, should set to be 0. When the state transition model is non-linear, generally there is no analytic expression of the result of the Maximum Likelihood estimation, so that the transition process should be converted into a discrete form to search all possible states.

\subsection{Sensing model}
\label{section:sensing}

Because all sensors in this work are activated by wireless power transmission, the power consumption becomes one of the most important determining factors in choosing the sensor.
Therefore, the sensors employed in this work are infrared directional binary human detection sensors. The sensors cannot directly give the information of user's location or moving status. The sensing outputs are binary only. The output 1 represents that the sensor detects the user in its coverage. And the output 0 represents that sensor does not detect any user in its coverage. Moreover, the infrared sensor is the cheapest solution among all types of sensors which can be used in localization, such as ultrasonic sensors and microwave sensors.

\begin{figure}[!t]
	\centering
	\includegraphics[width=7cm]{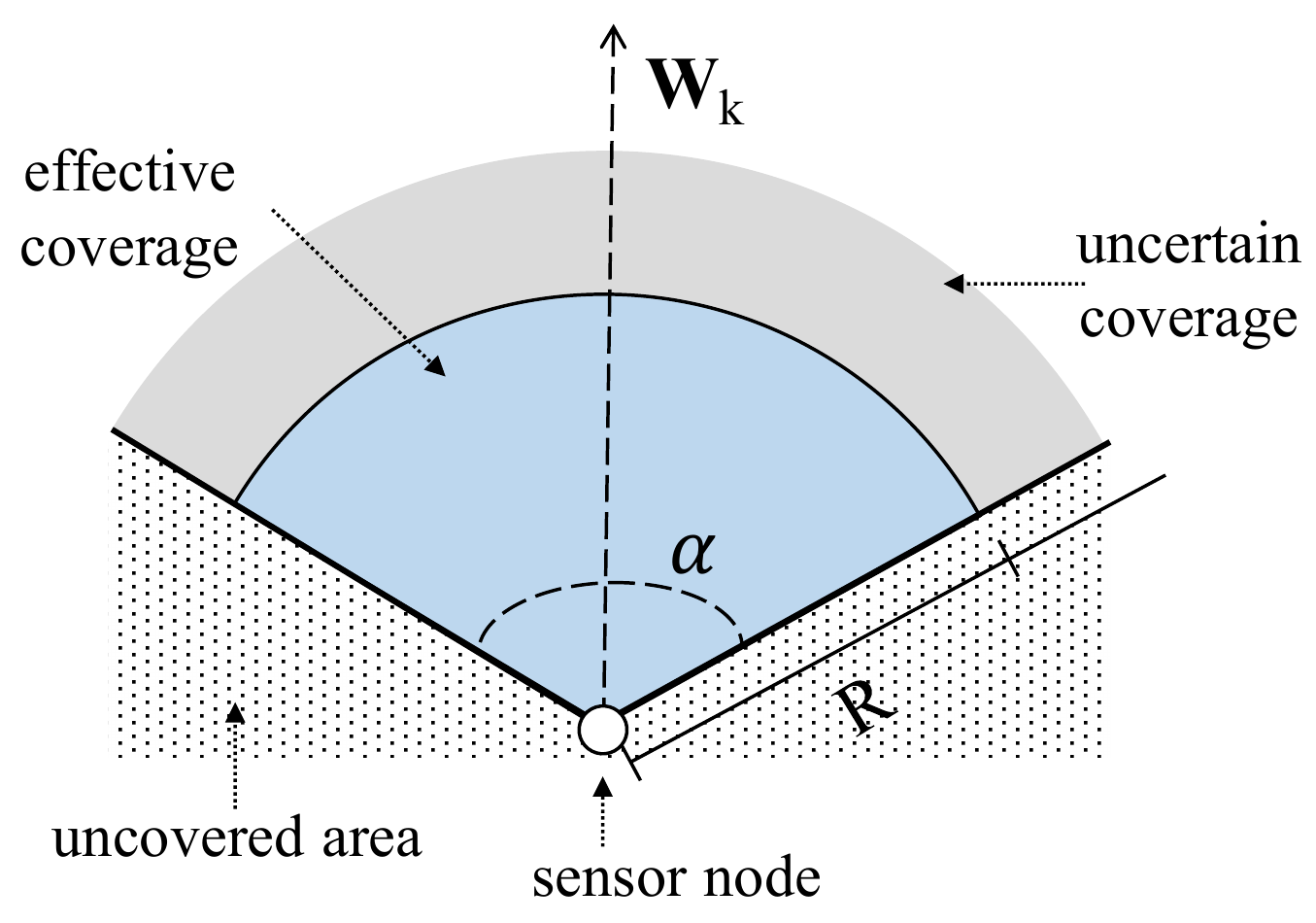}
	\caption{Directional sensing coverage model.}
	\label{fig:sensor}
\end{figure}

The employed sensors are directional sensors with fan-shaped sensing coverage, comparing with the disk-shaped coverage of omni-directional sensors. Considering the working mechanism of infrared sensors, distinct sensing-boundary in the fan-shape's two edges of radii are assumed. And we also assume that the detection probability is uniform within the effective coverage radius $R$, and when outside of effective coverage, it decreases exponentially with the distance. Fig.~\ref{fig:sensor} illustrates the directional coverage model. Therefore, the likelihood function of sensor detection is given by: 
\begin{equation}
	\label{eq:detection_likelihood}
	Pr(b_k | x_i,y_i)=
		\begin{cases}
			\begin{array}{*{20}{cl}}
				& P_{\rm{D}}c_{ik} &\text{ when}\   b_k=1\\ 
 				& 1-P_{\rm{D}}c_{ik} & \text{ when}\ b_k=0 
			\end{array} 	
		\end{cases}
\end{equation}
where 
\begin{equation}
c_{ik}=
	\begin{cases}
		\begin{array}{*{20}{cl}}
			0 &\text{ when}\ a_{ik}>\frac{\alpha}{2}\\
			e^{-\lambda \eta ^{\beta}} &\text{ when}\ R<d_{ik} \ \text{and}\  a_{ik}<\frac{\alpha}{2}\\
			1 & \text{ when}\ R\geq d_{ik}\  \text{and}\  a_{ik}<\frac{\alpha}{2}
		\end{array}
	\end{cases}
\end{equation}
where $b_k$ is the output sensing measurement of sensor $k$. $(x_i,y_i)$ denotes the position of object. $P_{\rm{D}}$ is the detection probability within the sensor's effective coverage radius $R$. $d_{ik}$ is the distance between sensor $k$ and position $(x_i,y_i)$. $a_{ik}$ is the angle between sensor's working direction $\bm{W}_k$ and the vector from sensor $k$ to position $(x_i,y_i)$. $\alpha$ is the view angle of the sensor. $c_{ik}$ is the attenuation factor of detection probability according to distance. $\eta = d_{ik}-R$. $\lambda$, $\beta$ are parameters shaping the attenuation of detection probability. 

Due to the mechanism that typically infrared sensor detects the fluctuation of infrared radiation impinging upon it from the objects in front of the sensor, the detection probabilities are different from the static and moving objects. 
By detecting the infrared radiation changes from the background and the moving object in its field of view, the sensor is sensitive to moving object, such as a walking human. While it is quite insensitive to static objects or small movements, because the tiny temperature changes may be treated as environment noises. This poses difficulties for the current infrared sensor based lighting control systems, because the user may stay somewhere even for a long time in an office.

And according to our test, the false alarm probability of the employed infrared sensors is negligible and can be assumed to be 0.

\subsection{User localization algorithm}
To estimate user's state, multi-sensor is deployed with overlapped coverage. Therefore, by combining the sensing measurements from multi-sensors, the large detection error of infrared human detection sensor can be overcome. In this section, a maximum likelihood estimation algorithm and a non-uniform resolution approximation algorithm are presented. The result will be used for LED light control in the next section.

User's state is estimated by maximizing the accumulated likelihood function that user is currently in certain state given the sensor outputs, which is given by: 
\begin{equation}
	\label{eq:maxlikehood}
	\argmax_{\bm{u_t}} L\left( \bm{u_t} | \bm{\mathcal{B}_t} \right)  = \argmax_{\bm{u_t}} Pr \left( \bm{\mathcal{B}_t}|\bm{u_t} \right)
\end{equation}
where $\bm{\mathcal{B}_t}=[ \bm{b_1},\ \bm{b_2},...,\bm{b_t}]$, and $\bm{b_t}=[ {b_{t,1}},\ {b_{t,2}},...,{b_{t,N}}]$ is the sensing vector of all sensors at time $t$, and $N$ is the number of sensors. All the sensing measurements are temporally and spatially conditionally independent. Therefore, for convenience of calculation, the accumulated likelihood can be separated into two parts: current likelihood and historical likelihood:
\begin{eqnarray}
	L\left( \bm{u_t} = \bm{s_i} | \bm{\mathcal{B}_t}\right)  &=& Pr \left( \bm{\mathcal{B}_t}|\bm{u_t}=\bm{s_i} \right)  \nonumber \\
	&=& Pr \left( \bm{b_t}|\bm{u_t}=\bm{s_i} \right) Pr \left( \bm{\mathcal{B}_{t-1}}|\bm{u_t}=\bm{s_i} \right) \nonumber \\
\end{eqnarray}
In our cases, the detection probability of employed sensor varies when the user is static (namely, $\bm{u_{t-1}}=\bm{s_{i}},\ \bm{u_t}=\bm{s_i}$) and moving (namely, $\bm{u_{t-1}}=\bm{s_{i}},\ \bm{u_t}=\bm{s_j},\ j \neq i$). Therefore, current likelihood is further separated into these two cases and calculated as:
\begin{eqnarray}
	\label{eq:current}
	&&Pr \left( \bm{b_t}|\bm{u_t}=\bm{s_i} \right) \nonumber \\
	&=& \sum_{j=1}^M{Pr \left( \bm{b_t}|\bm{u_{t-1}}=\bm{s_{j}}, \bm{u_{t}}=\bm{s_{i}} \right) T_{ij}} \nonumber \\
	&=& Pr \left( \bm{b_{t}} | \bm{u_{t-1}}=\bm{s_{i}}, \bm{u_t}=\bm{s_i} \right) T_{ii} \nonumber \\
	&+& \sum_{j=1, j \neq i}^M{Pr\left( \bm{b_t}|\bm{u_{t-1}}=\bm{s_j}, \bm{u_t}=\bm{s_i}\right) T_{ij} }
\end{eqnarray}
where $M=|\bm{\mathcal{S}}|$ denotes the number of states in state space. 
$T_{ii}$ and $T_{ij}$ are the transition probability between two successive states for a static and moving status correspondingly. They are defined by the movement model and can be derived from Eq.~\eqref{eq:transition}. 

The historical likelihood can be calculated as:
\begin{eqnarray}
	\label{eq:history}
	&&Pr \left( \bm{\mathcal{B}_{t-1}} | \bm{u_{t}} = \bm{s_i} \right) \nonumber \\
	&=& \sum_{j=1}^{M} { Pr \left( \bm{\mathcal{B}_{t-1}} | \bm{u_{t-1}} = \bm{s_{j}} \right) T_{ij} } \nonumber \\
	&=& \sum_{j=1}^{M} { L \left( \bm{u_{t-1}}=\bm{s_{j}}| \bm{\mathcal{B}_{t-1}} \right) T_{ij} }
\end{eqnarray}
where $L( \bm{u_{t-1}}=\bm{s_{j}}| \bm{\mathcal{B}_{t-1}})$ is the accumulated likelihood in the previous time step $t-1$.

\begin{figure}[!t]
	\centering
	\includegraphics[width=8cm]{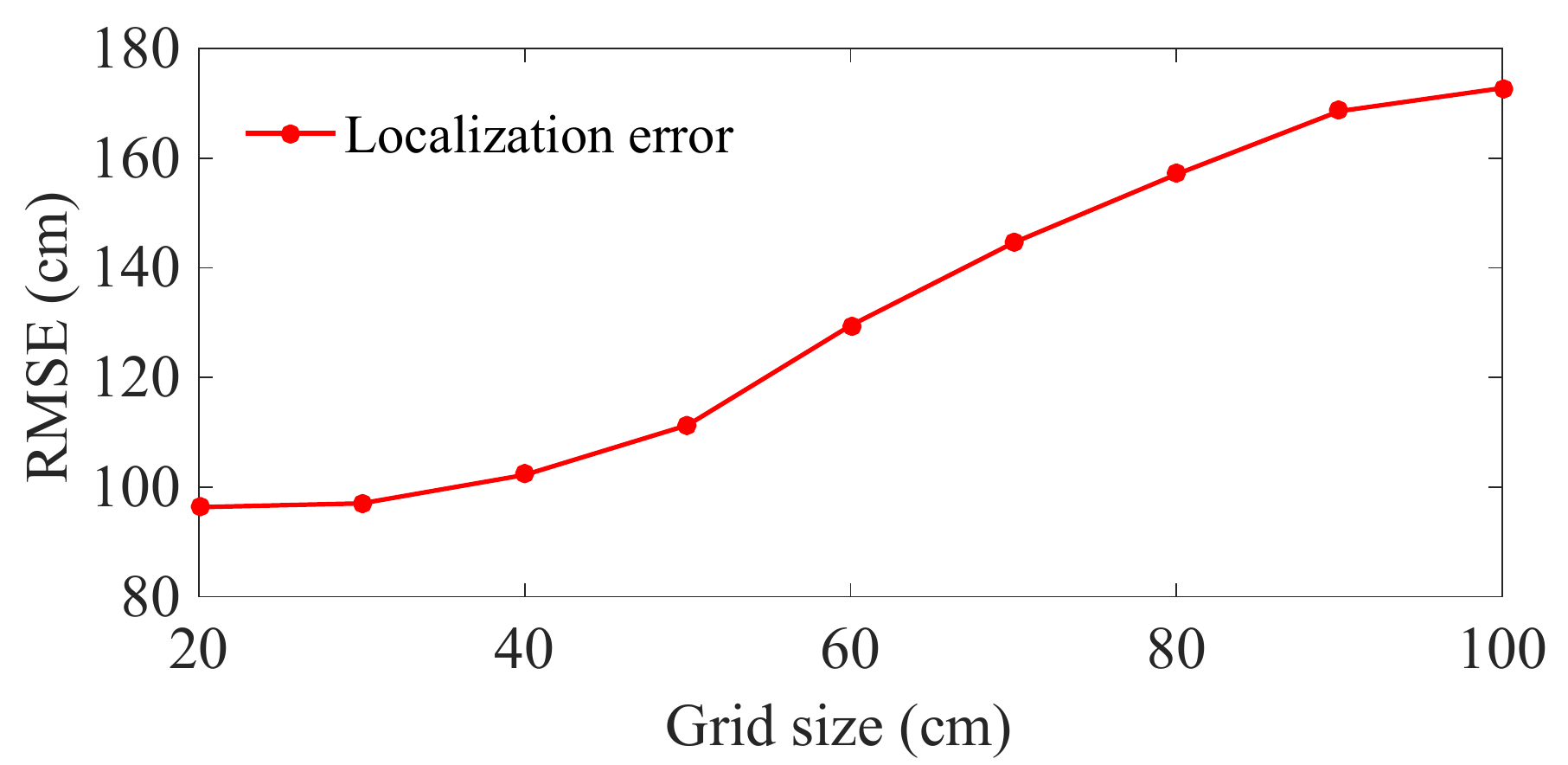}
	\caption{Localization error.}
	\label{fig:locerror}
\end{figure} 

Generally, it is difficult to obtain the closed form of result for Eq.~\eqref{eq:maxlikehood}. The numerical calculation should be done on discrete states in the state space. The estimation error is inevitable because of the grid search and the sensing capability of the employed directional infrared binary human detection sensors. Figure~\ref{fig:locerror} shows the relationship between grid size and the localization error.

Therefore, to compensate the estimation error and avoid loss of user satisfaction, not only the position with maximum likelihood, but the illumination requirement in all the positions in a set $G_{\rm{T}}$ should be satisfied. The set of location candidates $G_{\rm{T}}$ contains positions whose current total likelihood is larger than threshold $\rm{Th_c}$. 
\begin{equation}
	G_{\rm{T}}=\{ (x_i,y_i)|L\left( \bm{u_t} = \bm{s_i} | \bm{\mathcal{B}_t} \right)> {\rm{Th_c}}, \bm{s_i}\in \bm{{S}} \} 
\end{equation}

Apparently the setting of $\rm{Th_c}$ affects the performance of LED lights control and should balance the power consumption and user's satisfaction. 
Currently in the verification experiments, $\rm{Th_c}$ is set empirically. 
The optimized threshold should depend on locations, because of the location-invariant estimation accuracy, which is caused by the non-uniform deployment of directional sensors and user's nonuniform moving patterns in a restricted indoor space. And the illumination distribution resulted from LED lights' spatial distribution should also be taken into consideration. The theoretical analyses of optimized thresholds will be included in the future work. 

\begin{figure}[!t]
	\centering
	\subfigure[Top view diagram of target space.]{
		\label{fig.sub.topview}
		\includegraphics[height=5.8cm,width=8.6cm]{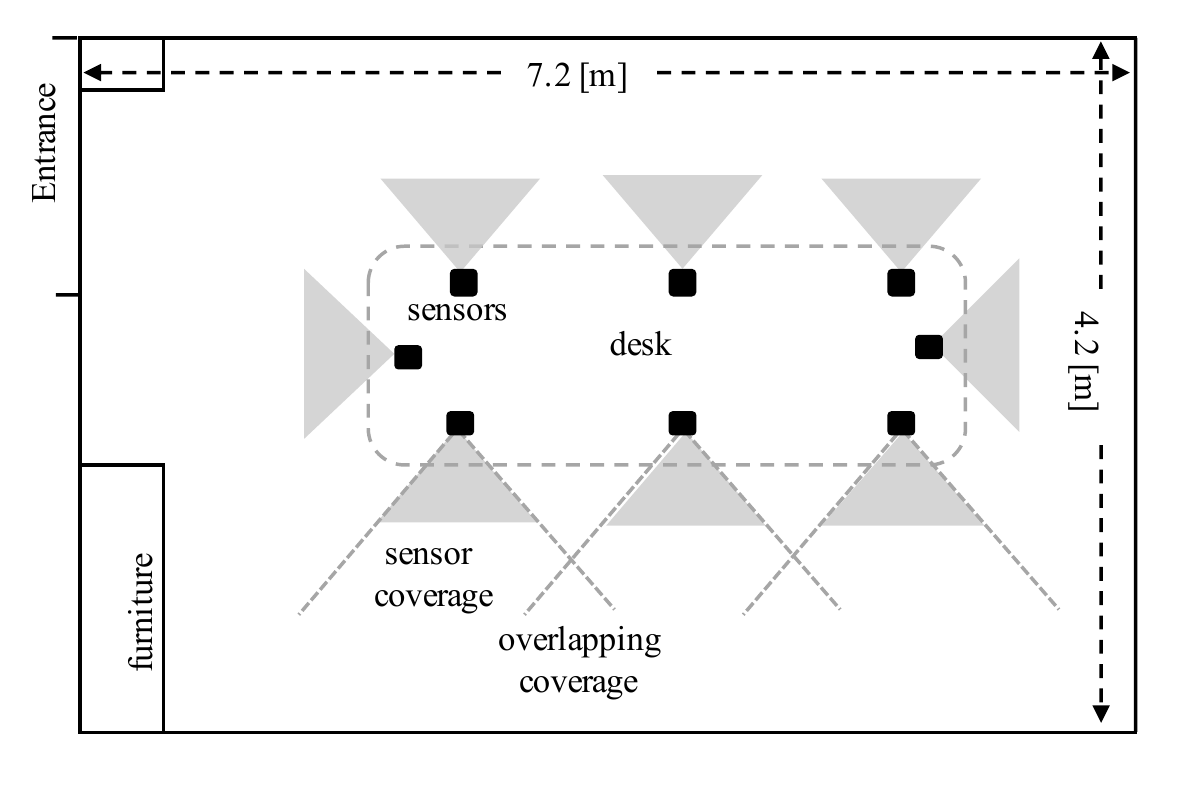}
	}
	\subfigure[Experiments results.]{
		\label{fig.sub.results}
		\includegraphics[height=5.8cm,width=8.6cm]{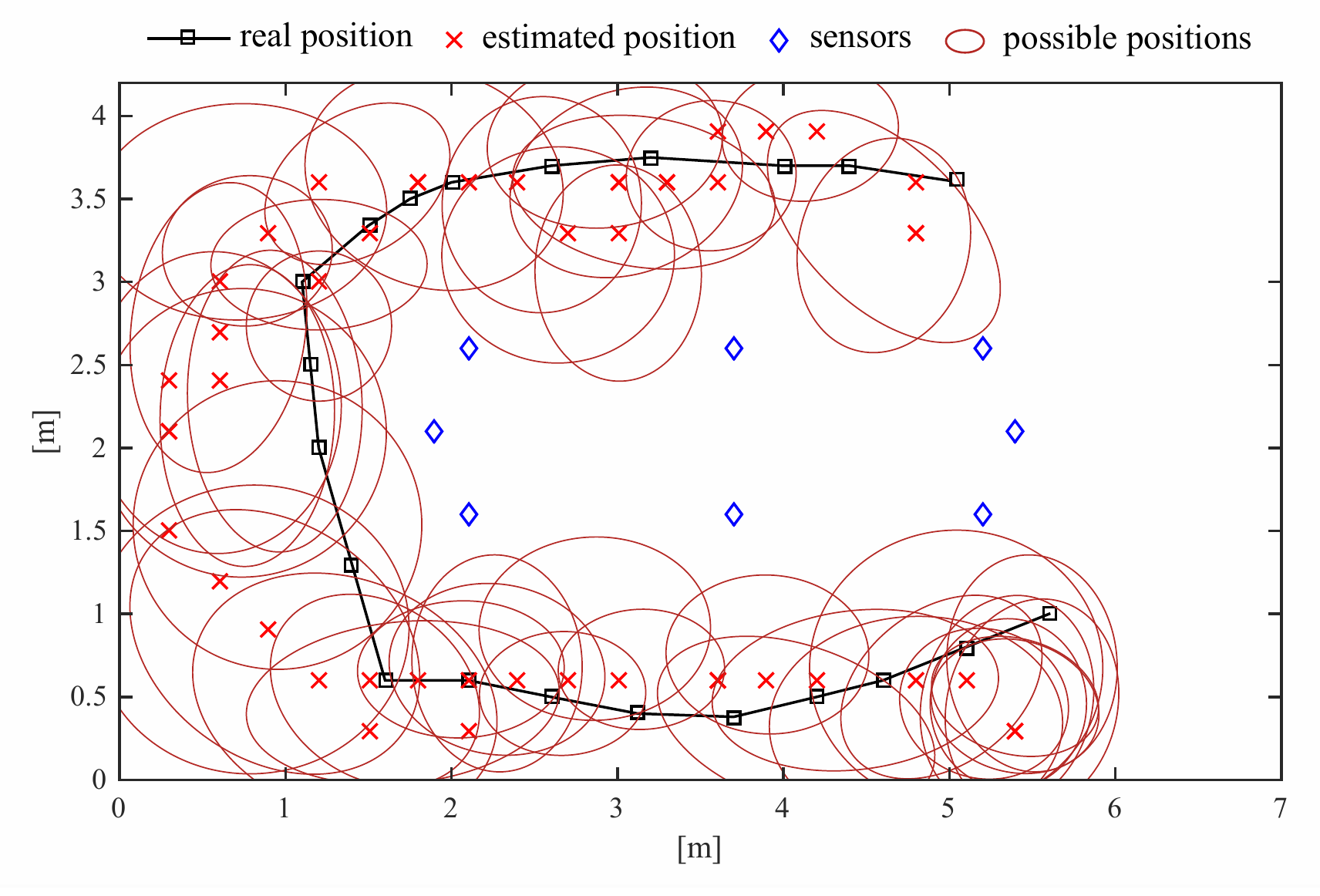}
	}
	\caption{Results of user localization experiments.}
	\label{fig:locationresults}
\end{figure}

\subsection{User localization experiment}
An experiment is conducted to verify the performance of the battery-less human detection sensor network. Because the proposed lighting control system is mainly designed for such indoor environments as office, the target space is a typical indoor office environment. It includes desks and pathways, and the size is 7.2~m $\times$ 4.2~m, as shown in Fig.~\ref{fig.sub.topview}. 
Human detection sensors are placed on the desks. 
In the experiment, a user walks around in the space, from top right corner to down right corner. 

The real positions and experimental results are shown in \ref{fig.sub.results}. 
The black line denotes user's real routes, and the red x denotes the position with maximum likelihood. 
The ellipses indicate the area of all possible positions, which are positions with probabilities higher than a threshold, as described in Sec~\ref{sec:localization}.
Although the sensing capacity of a single binary infrared sensor is very coarse and cannot directly give the information of user's position, but by deploying a series of distributed sensors, the estimated positions are close to real routes. The performance is also related to user's moving status, the number and placement of sensors. Additionally, although it is a single user experiment, it is easy to extend to multi-user cases if the number of users is known. 
With the consideration of calculation amount and to ensure the accuracy of LED control, the space is divided into grids whose size is $30cm \times 30cm$. And it results in the localization error of 91 cm. Note that the real positions of the user are derived from the video of the experiment, so there could exist difference between the localization error in experiment and theoretically in Fig.~\ref{fig:locerror}.  

\section{Lighting control experiment} 
\label{sec:experiment}
\begin{figure}[!t]
	\centering
	\includegraphics[width=8.6cm]{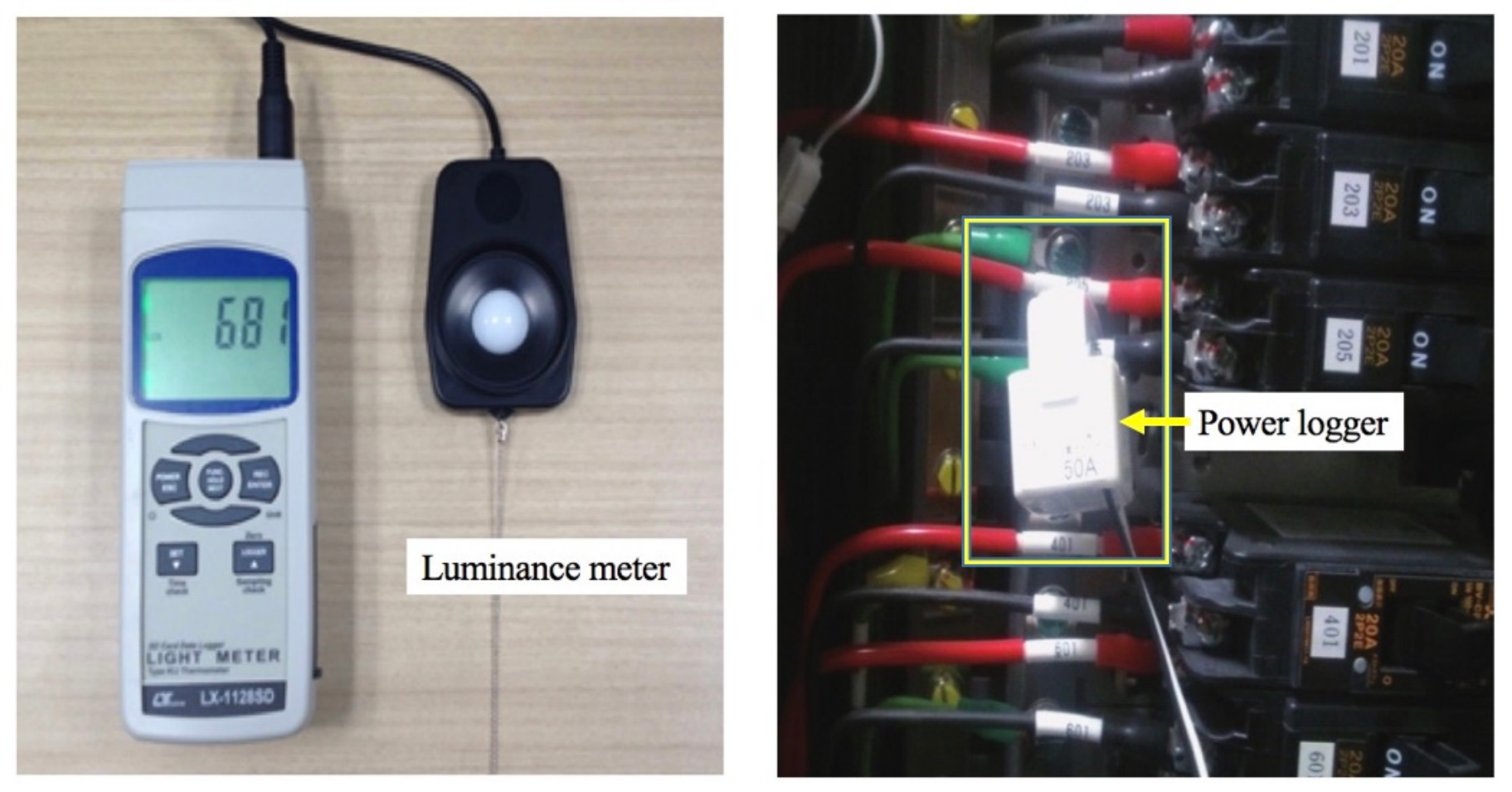}
	\caption{Experiment devices.}
	\label{fig:meters}
\end{figure}

The LED light control system is implemented in an indoor office environment. To evaluate the performance, an experiment is conducted. The test user walks in the office. Meanwhile the user measures and records the illumination by a handheld illumination meter at working surface height. And at the same time, the real-time lighting power consumption of the office is recorded by a power logger, which is installed in the cable box and clipped on the electricity cable. The illumination meter and power logger are shown in Fig.~\ref{fig:meters}. 

\subsection{Experiment environment}

The experiment area is a part of the office. Seven LED lights are under the ON/OFF control by the controller. Wireless power transmitters are embedded in all LED lights. Eight battery-less human detection sensors are put on the desk to detect user's location and track user's motion. By our measurements, practical false alarm probability of the employed sensor is approximately 0, and for moving user detection probability is about 0.8, while for a static user it is only 0.1 which is quite low and makes it almost impossible to effectively detect the user. The reason is discussed in Sec.~\ref{section:sensing}. 
And the minimum required illumination is set to be 400 lux according to the standard of indoor lighting levels of Japan \cite{jis}. 

The full-on working power consumption of an LED light is 100~W. Each wireless power transmitter transmits power of 1~W, which is quite small compared to the power consumption of LED light and can be further decreased by optimization of energy transmission.

\begin{figure}[!t]
	\centering
	\includegraphics[width=8.5cm]{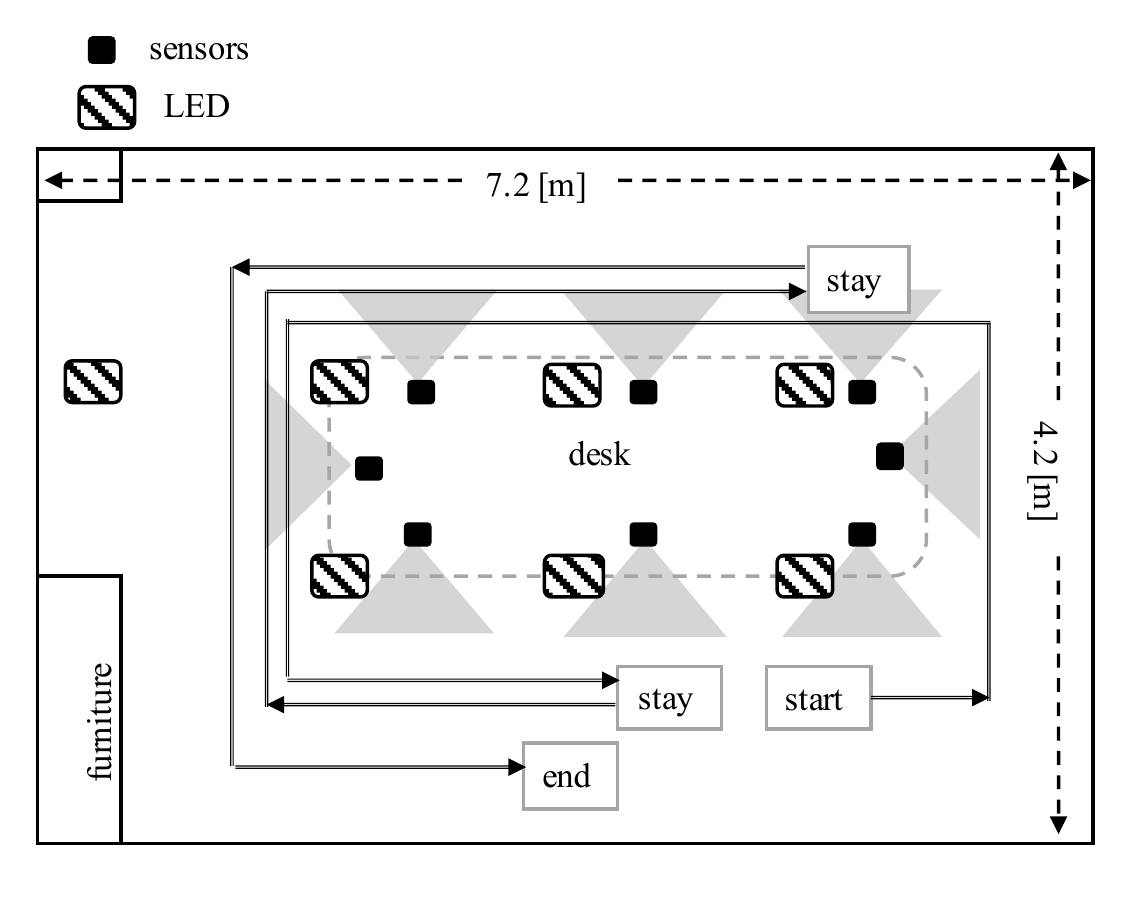}
	\caption{Experiment environment and walking path.}
	\label{fig:path}
\end{figure}

The walking path is as shown in Fig.~\ref{fig:path}. To simulate the real office user, the test user also switches between moving mode and static mode. If the system works well, only the lights near user's location and path will be turned on, while others will be turned off to save the power.

\subsection{Experimental result} 

The performance of proposed scheme is compared with the individual control, batch control and the ideal case of perfect user localization. 
In individual control, each LED light is individually triggered by a single sensor which is embedded beside LED light on the ceiling and can cover the area below the LED light. 
In the batch control, which is currently the most widely used sensor-based lighting control scheme, if any of the sensors detects user's presence, the controller switches on all LED lights simultaneously, and if all sensors detect absence for a given delay period, it switches off all the LED lights.

\begin{figure}[!t]
\centering
\includegraphics[width=8.6cm]{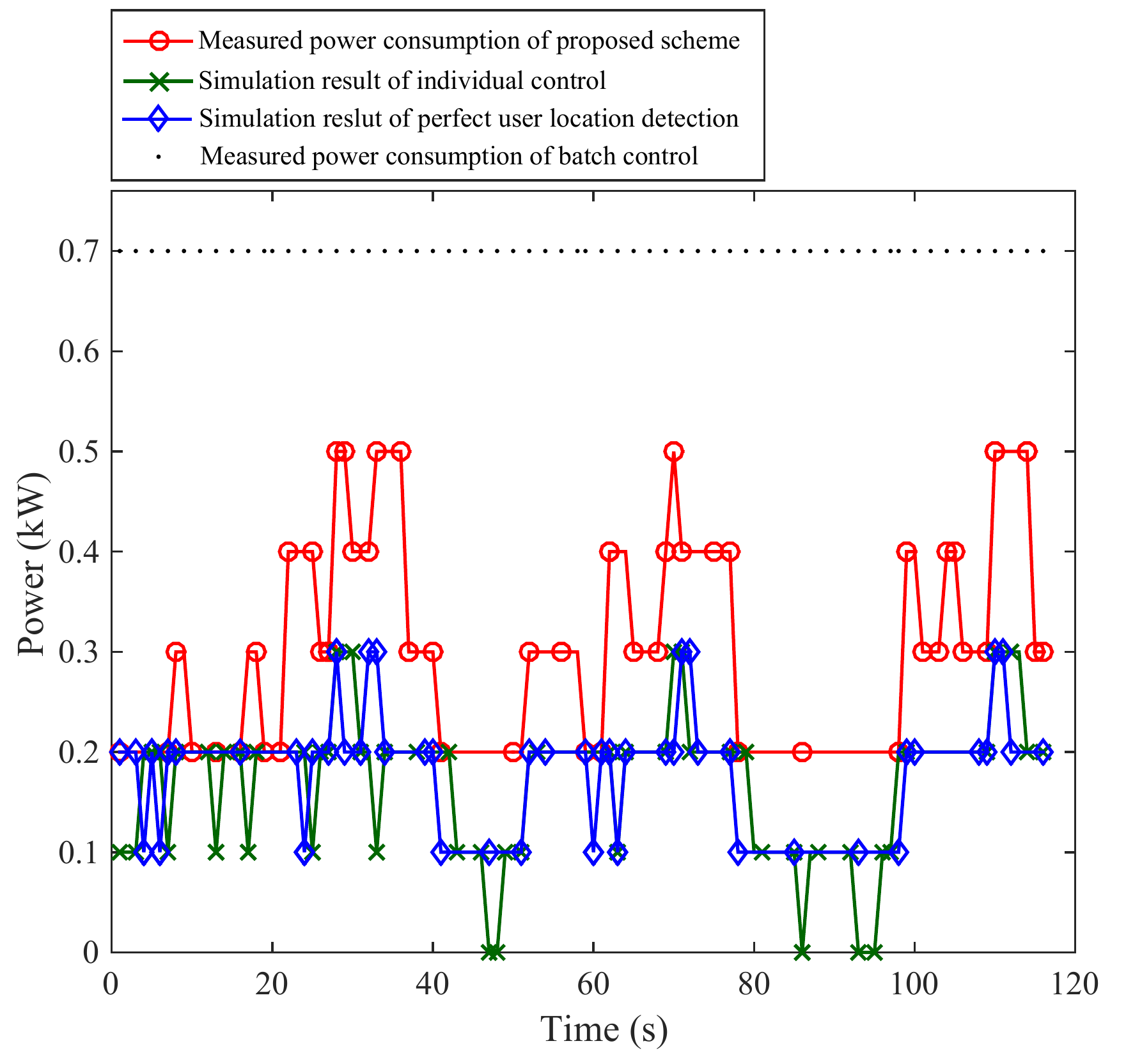}
\caption{Experimental result of power consumption.}
\label{fig:result2}
\end{figure}

Experimental results of power consumption are given in Fig.~\ref{fig:result2}. 
The red curve is the measured results using the proposed scheme. The green curve is the simulation results of individual control. The blue curve is the simulation results of the case of perfect user localization. And the black dot line drawn at the power of 0.7 kW is the power consumption for the batch control case.

\begin{figure}[!t]
	\centering
	\includegraphics[width=8.8cm]{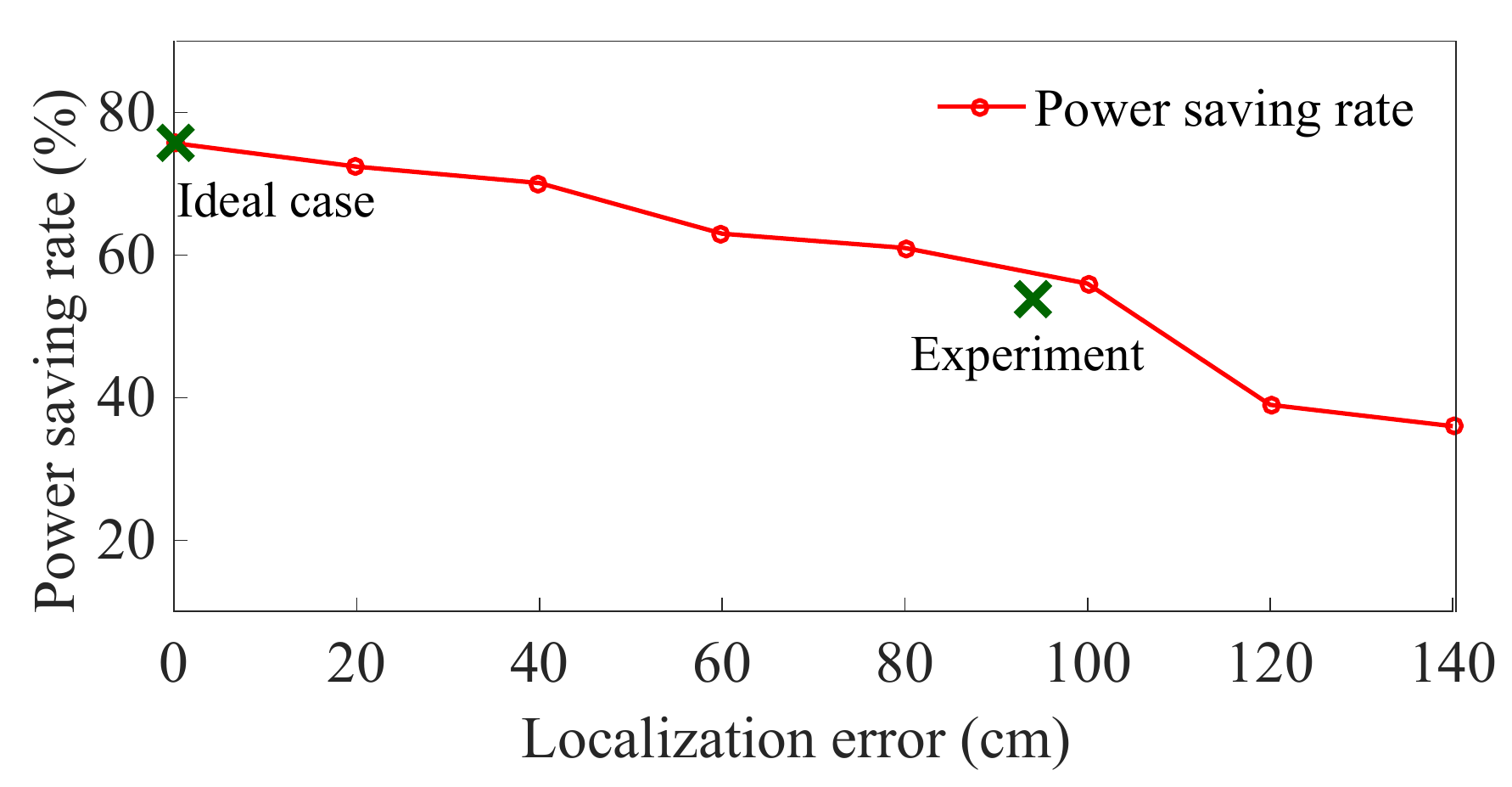}
	\caption{Power saving rate.}
	\label{fig:savingpower}
\end{figure}

In our proposed LED control scheme, in most cases, only 2 or 3 working LED lights are enough to satisfy user's illumination requirement, compared to that in the batch control scheme all 7 LED lights are on. Therefore, the power consumption of our LED light control system is always lower than that of batch control. In the experiment, its power consumption is 57\% less than the batch control. The power consumed by the LED control system is almost negligible compared to the working power of LED light and saved power. However, the practical power consumption is still about 1.7 times of ideal case. 
This is caused by the inaccurate estimation of user's location. As discussed in Sec.~4.3, in this case, the controller has to satisfy the illumination requirement not only in the estimated position, which is the position with maximum possibility, but also the potential positions, which is the positions with possibilities larger than a threshold.
Thus obviously the energy saving performance of proposed lighting control scheme highly depends on the information of user's position. Figure~\ref{fig:savingpower} shows the effect of localization error on the energy saving performance from simulation and experiment.
Therefore, the performance could be improved by more accurate indoor localization technique. And in the case of individual control, although the power consumption is the lowest, it sacrifices the user satisfaction because it fails to switch the corresponding LED lights on correctly sometimes due to sensors' miss detection.

\begin{figure}[!t]
\centering
\includegraphics[width=8.6cm]{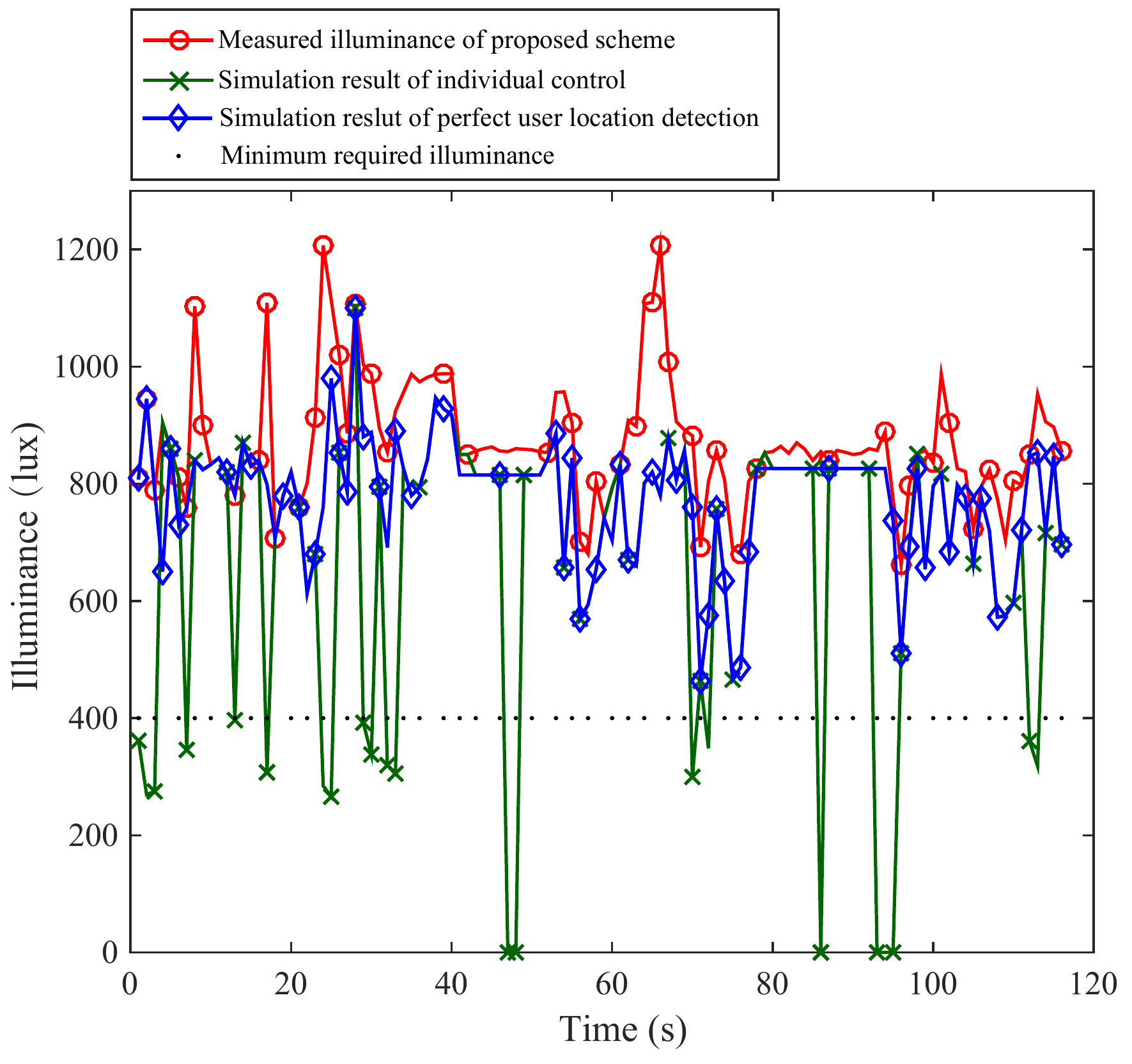}
\caption{Experimental result of illumination.}
\label{fig:result1}
\end{figure}

Figure~\ref{fig:result1} gives the experimental results of illumination. 
The red, green and blue curves correspond to the power consumption curves with the same color in Fig.~\ref{fig:result2}.
The red curve is the measured illumination results using the proposed scheme. And the green curve is the illumination results of the simulation, in which each LED light is individually triggered by a single sensor. The blue curve is the simulation results, in which user localization is perfectly detected. 
And the black dot line drawn at the illumination of 400 lux is the minimum required illumination level.

During the experiment, the practically measured illumination is always higher than 400~lux with 100\% probability, i.e., the user's illumination requirement can always be satisfied. 
The average illumination is 867 lux, the maximum illumination intensity is 1207 lux and the minimum illumination intensity is 662 lux. However, Fig.~\ref{fig:result1} shows that even the minimum practical illumination is larger than the required illumination, which is pre-set to be 400 according to the standard of indoor lighting levels of Japan. This is because in the current stage only ON/OFF LED light control is conducted.
In the case of dimmer control, the results should be improved. The illumination of individual control sometimes drops lower than 400~lux and even to 0~lux, due to the miss detection of infrared sensors, especially for a static user.

Additionally, the measured illumination curve in Fig.~\ref{fig:result1} seems to be fluctuating severely when the test user is moving, but in practice, such fluctuation is generally imperceptible for users with the naked eyes, especially when walking. On the other hand, in the case that the testing user is staying somewhere in the room to simulate daily working status, the number of working LED lights does not change and hence the illumination is quite stable.

\section{Conclusion}
\label{sec:conclusion}
In this work, we designed an LED light control system based on user's location using a battery-less sensor network, and implemented it in a practical indoor office environment. 
To address the challenge of energy stability, all sensors were battery-less and activated by multiple wireless energy transmitters embedded in LED ceiling lights.
We deployed a distributed set of sensors to overcome the detection errors of infrared sensors. By processing the output of the multiple sensors, user location and motion are detected, tracked and predicted by a maximum likelihood algorithm. 
LED lights' switch states were controlled based on the user location detection and pre-set minimum required illumination. 
The experimental result shows that this LED light control system reduces the energy consumption significantly, 57\%, compared to the batch control scheme, and satisfies user’s illumination requirement with 100\% probability.

\section*{Acknowledgement}
This work is partially supported by Grants-in-Aid for Scientific Research from Japan Society for the Promotion of Science with project number of 24246069.

\profile[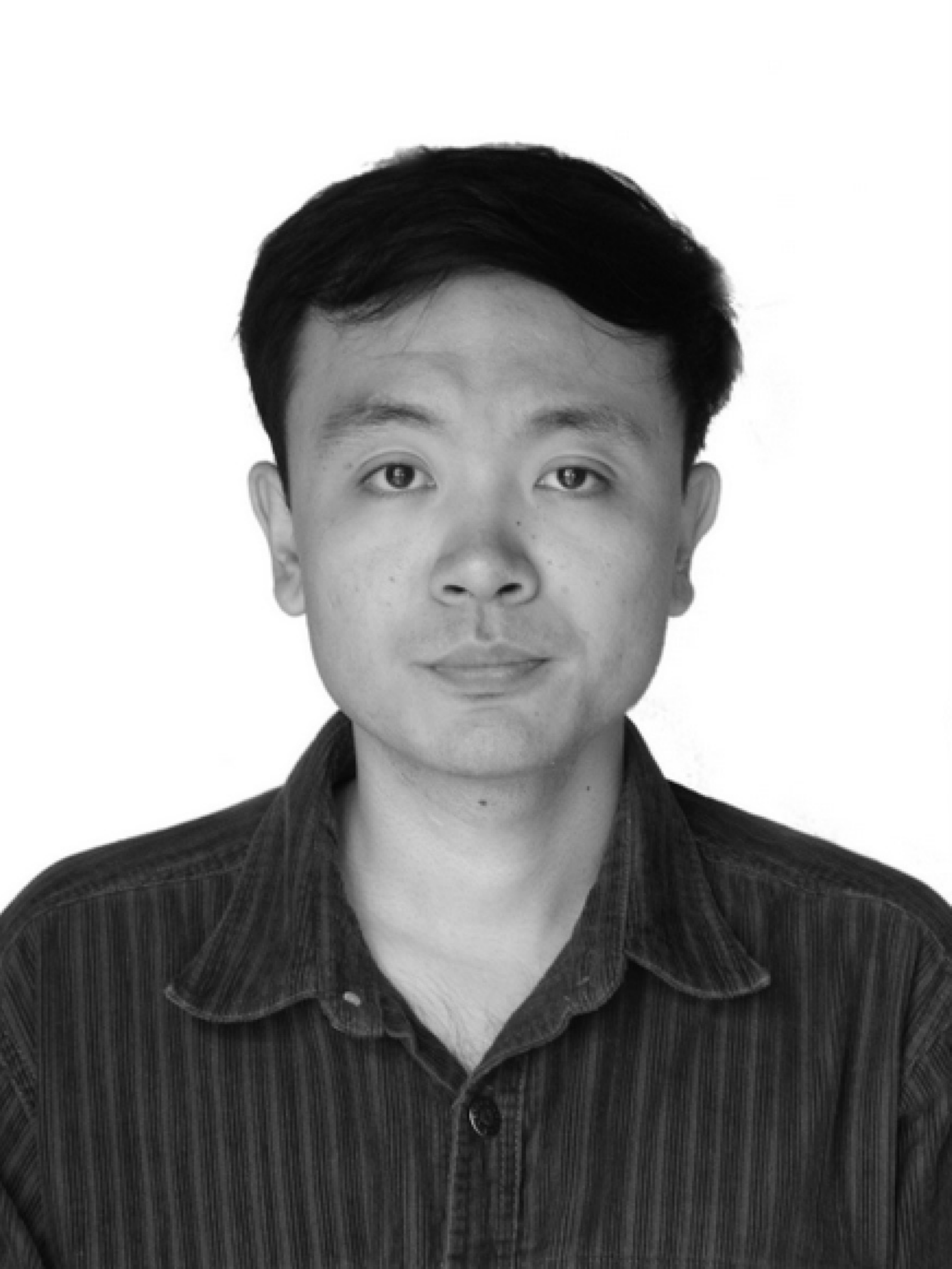]{Tao Yu}
{
received the B.E. degree in communication engineering from Taiyuan Institute of Technology, China, in 2008, and the M.E. degree in signal and information processing from Communication University of China, in 2010. He is currently working toward the Ph.D. degree in the department of Electrical and Electronic Engineering, Tokyo Institute of Technology. His research interests are building energy management and sensor networks.}

\profile[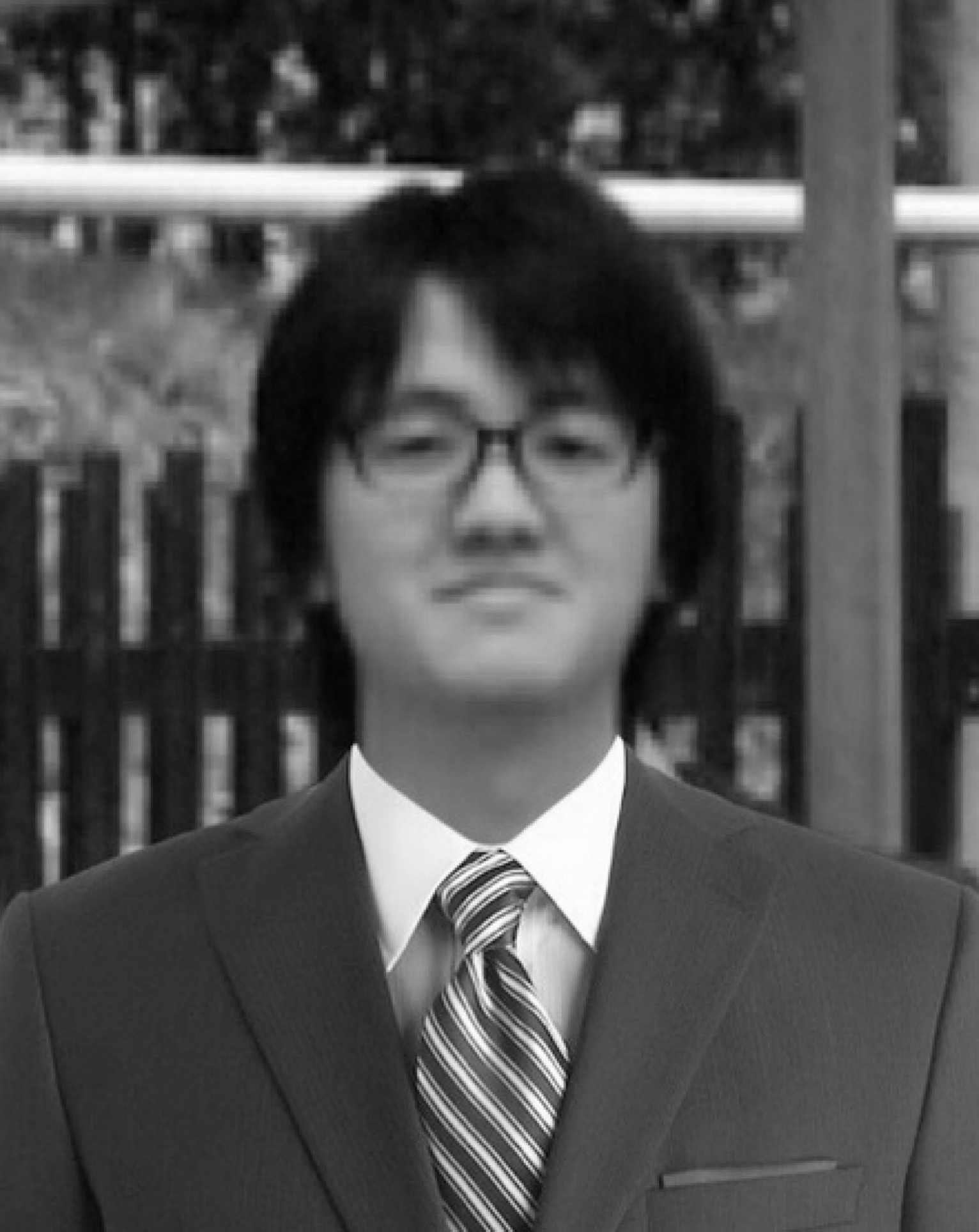]{Kuki Yusuke}
{
received the B.E. degree in electronic and information engineering from Osaka University, Osaka, Japan in 2013, and the M.E. degree in electrical, electronic and information engineering from Osaka University, Osaka, Japan in 2016.}

\profile[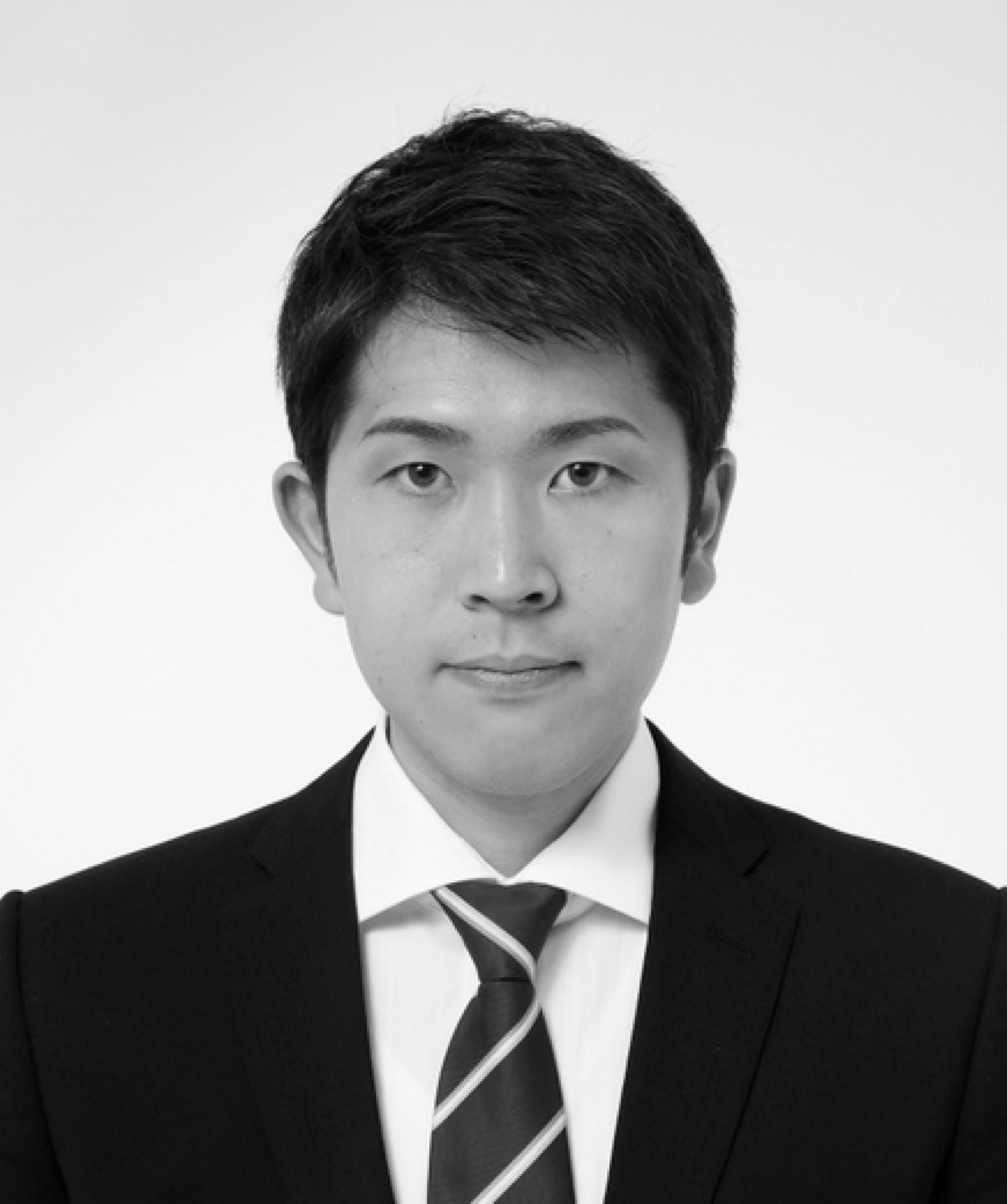]{Gento Matsushita}
{
received the B.E. degree in electronic and information engineering from Osaka University, Osaka, Japan in 2014, and the M.E. degree in electrical, electronic and information engineering from Osaka University, Osaka, Japan in 2016, where he conducted the research of wireless energy transmission, wireless sensor network systems and low-power circuit design.}

\profile[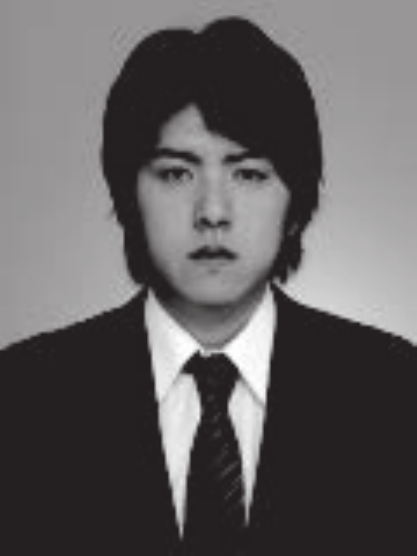]{Daiki Maehara}
{
received the B.E. and M.E. degrees in electrical and electronic engineering from Tokyo Institute of Technology, Japan, in 2011 and 2013 respectively. He is currently a Ph.D. student at Tokyo Institute of Technology. He received the Best Paper Award from both IEICE SR technical committee and IEICE Communication Society in 2013 and 2014 respectively. His research interests are wireless energy transmission, RFID, sensor network systems and low-power circuit design.
}

\profile[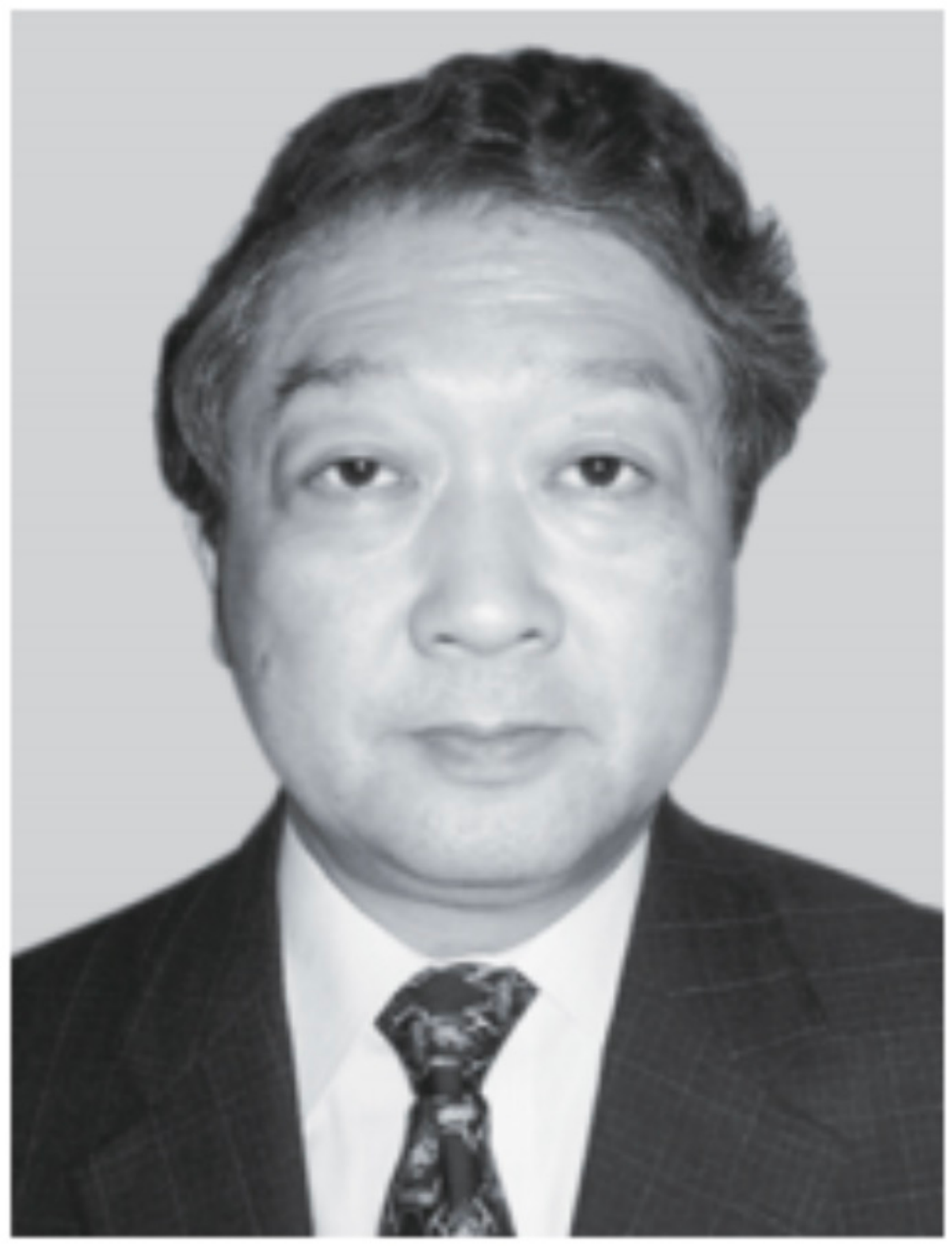]{Seiichi Sampei}
{
received the B.E., M.E., and Ph.D degrees in electrical engineering from Tokyo Institute of Technology, Japan, in 1980, 1982 and 1991, respectively. From 1982 to 1993, he was with the Communications Research Laboratory, Ministry of Posts and Telecommunications. During 1991 to 1992, he was at the University of California, Davis, as a visiting researcher. In 1993, he joined the Faculty of Engineering, Osaka University, and he is currently a Professor in the department of Information and Communications Technology, Osaka University. He has developed active modulation, intelligent radio transmission/access, cognitive wireless networking, and wireless distributed network techniques. He received the Shinohara Young Engineering Award, the Achievements Award and Communications Society Best Paper Award from the IEICE, the Telecom System Technology Award from the Telecommunications Advancement Foundation, the DoCoMo Mobile Science Award from the Mobile Communication Fund, and the Ericsson Telecommunications Award. He is a member of the Institute of Image Information and Television Engineers(ITE) and a Fellow of IEEE.
}

\profile[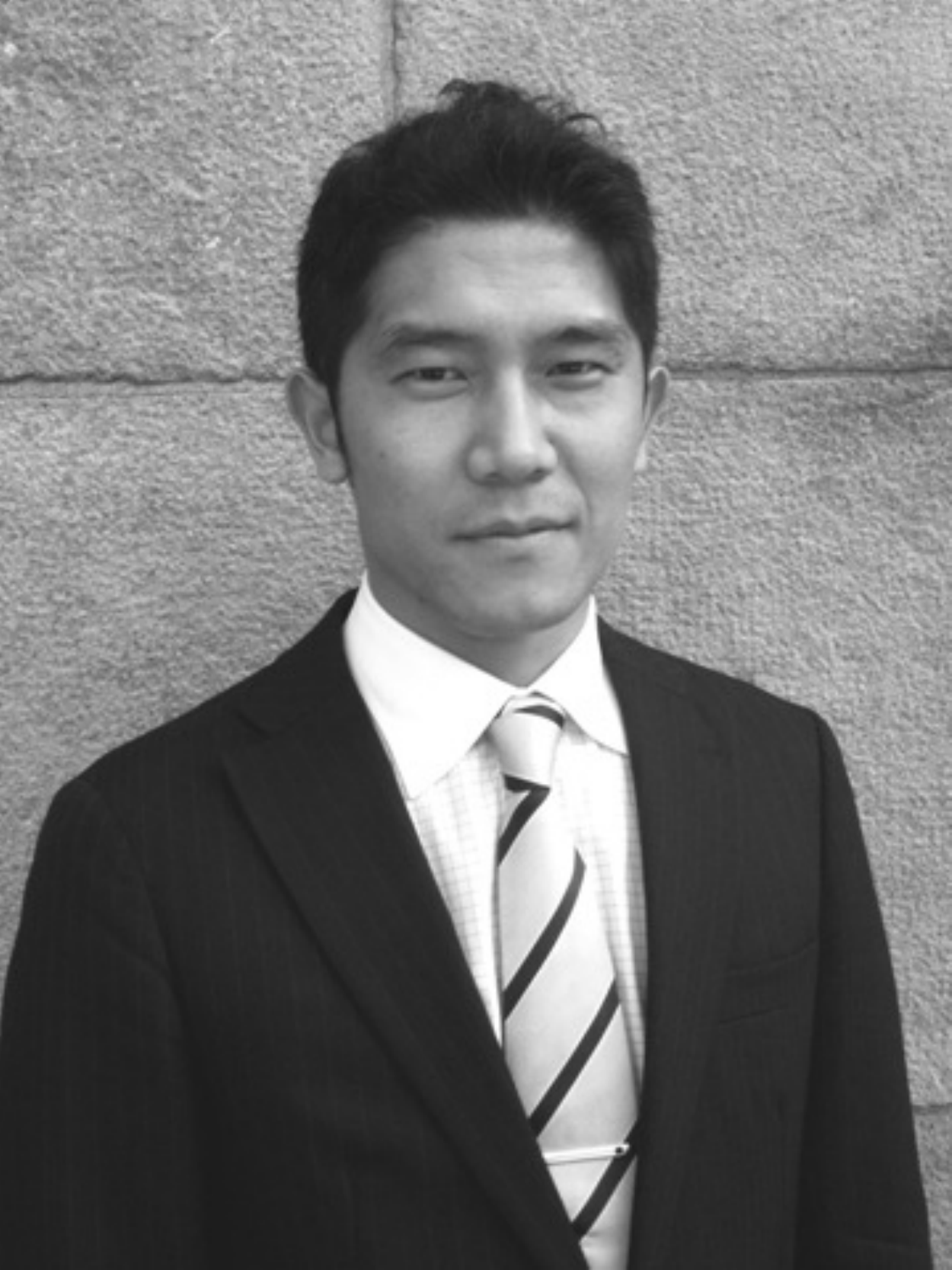]{Kei Sakaguchi}
{
Kei Sakaguchi received the M.E. degree in Information Processing from Tokyo Institute Technology in 1998, and the Ph.D degree in Electrical \& Electronics Engineering from Tokyo Institute Technology in 2006. Currently, he is working at Tokyo Institute of Technology in Japan as a Professor and at the same time he is a Senior Scientist at Fraunhofer HHI in Germany. He received the Outstanding Paper Awards from SDR Forum and IEICE in 2004 and 2005 respectively, and three Best Paper Awards from IEICE communication society in 2012, 2013, and 2015. He also received the Tutorial Paper Award from IEICE communication society in 2006. He served as a TPC co-chair in the IEEE 5G Summit in 2016, a General co-chair in the IEEE WDN-5G in 2017, and a Industrial Workshop co-chair in the IEEE Globecom in 2017. His current research interests are in 5G cellular networks, millimeter-wave communications, and wireless energy transmission. He is a member of IEICE and IEEE.
}

\end{document}